\documentclass[preprint2]{aastex}

\newcommand{\be}{\begin{equation}}
\newcommand{\ee}{\end{equation}}
\newcommand{\bea}{\begin{eqnarray}}
\newcommand{\eea}{\end{eqnarray}}
\newcommand{\gtabouteq}{\,\hbox{\raise 0.5 ex \hbox{$>$}\kern-.77em 
                    \lower 0.5 ex \hbox{$\sim$}$\,$}}       
\newcommand{\ltabouteq}{\,\hbox{\raise 0.5 ex \hbox{$<$}\kern-.77em 
                     \lower 0.5 ex \hbox{$\sim$}$\,$}}

\shorttitle{N4845}
\shortauthors{Irwin et al.}

\begin{document}


\title{CHANG-ES V: \\
Nuclear Radio Outflow in a Virgo
Cluster Spiral after a Tidal Disruption Event
}


\author{
Judith A. Irwin\altaffilmark{1},
Richard N. Henriksen\altaffilmark{1},
Marita Krause\altaffilmark{2},
Q. Daniel Wang\altaffilmark{3},
Theresa Wiegert\altaffilmark{1},
Eric J. Murphy\altaffilmark{4},
George Heald\altaffilmark{5},
and
Eric Perlman\altaffilmark{6}
}

\altaffiltext{1}{Dept. of Physics, Engineering Physics, \& Astronomy, Queen's University,
    Kingston, Ontario, Canada, K7L 3N6 {\tt irwin@astro.queensu.ca, henriksn@astro.queensu.ca, twiegert@astro.queensu.ca }.}
\altaffiltext{2}{Max-Planck-Institut f{\"u}r Radioastronomie,  Auf dem H{\"u}gel 69,
53121, Bonn, Germany,
{\tt mkrause@mpifr-bonn.mpg.de}.} 
\altaffiltext{3}{Dept. of Astronomy, University of Massachusetts, 710 North
Pleasant St., Amherst, MA, 01003, USA, 
{\tt wqd@astro.umass.edu}.}
\altaffiltext{4}{US Planck Data Center, The California Institute of Technology, MC 220-6, Pasadena,
CA, 91125, USA, {\tt emurphy@ipac.caltech.edu}.}
\altaffiltext{5}{Netherlands Institute for Radio Astronomy (ASTRON), 
Postbus 2, 7990 AA, Dwingeloo, The Netherlands,
{\tt heald@astron.nl}.}
\altaffiltext{6}{Physics and Space Sciences Dept., Florida Institute of Technology,
150 West University Boulevard, Melbourne, FL, 32901, USA,
{\tt eperlman@fit.edu}.}




\begin{abstract}

We have observed the Virgo Cluster spiral galaxy, NGC~4845, at 1.6 and 6 GHz using the
Karl G. Jansky Very Large Array, as part of
 the `Continuum Halos in Nearby Galaxies -- an EVLA
Survey' (CHANG-ES).  The source consists of a bright unresolved core with
a surrounding weak central disk (1.8 kpc diameter). The core 
is variable over the 6 month time scale
of the CHANG-ES data and has increased by a factor of $\approx$ 6 since 1995. 
The wide bandwidths of CHANG-ES have allowed us to
determine the spectral evolution of this core which peaks {\it between} 1.6 and 6 GHz
(it is a GigaHertz-peaked spectrum source).
We show that the spectral turnover is dominated by synchrotron self-absorption and that
the spectral evolution can be explained 
by adiabatic expansion (outflow), likely in the form of a jet or cone.
The CHANG-ES observations serendipitously overlap in time with the hard X-ray
light curve obtained by Nikolajuk \& Walter
(2013) which they interpret as due to a tidal disruption event (TDE) of a
super-Jupiter mass object around a $10^5\, M_\odot$ black hole. 
We outline
a standard jet model, provide an explanation for the observed circular polarization,
and quantitatively suggest a link between the peak radio and peak X-ray emission via inverse Compton upscattering of the photons emitted by
the relativistic electrons.
We predict that it should be possible to resolve a young radio jet via VLBI 
as a result of this nearby TDE. 

\end{abstract}





\section{Introduction}
\label{sec:introduction}

The discovery of a hard X-ray source
at the center of the galaxy, NGC~4845,
by INTEGRAL (IGRJ12580+0134) 
 has been interpreted as the tidal
disruption of a super-Jupiter by a massive black hole \citep{NW13}.
As part of the Continuum Halos in Nearby Galaxies -- an EVLA\footnote{The Expanded Very Large Array is now
known as the Jansky Very Large Array.}
Survey (CHANG-ES), we have detected
a variable radio source (a compact core) in NGC4845 (Table~\ref{table:galaxy_parameters}), 
showing unambiguously that this galaxy harbours an active galactic nucleus (AGN).
The peak of
the X-ray light curve occurred on January 22, 2011. Our radio observations were carried out
approximately one year later (Table~\ref{table:image_parameters}) and overlap with
the time period of their light curve (their Fig.~8).

\begin{table}[h]
\begin{center}
\caption{Galaxy parameters\label{table:galaxy_parameters}}
\begin{tabular}{lc}
\tableline\tableline
{Parameter}  & {Value} \\ 
\tableline
Distance (Mpc)\tablenotemark{a} & 17  \\
Inclination (deg)\tablenotemark{b} & 81 \\
$V_\odot$ (km s$^{-1}$)\tablenotemark{c}& 1098 \\
$\Delta\,V$ (km s$^{-1}$)\tablenotemark{d} & $374 \pm\,0.4$    \\
HI mass ($M_\odot$)\tablenotemark{e} & $(2.1\,\pm\,0.4)\,\times\,10^8$ \\
Total mass ($M_\odot$)\tablenotemark{f} & $9.9\,\times\,10^{10}$\\
$I_{CO}$ (K km s$^{-1}$)\tablenotemark{g} & 6.07 \\
\tableline
\end{tabular}
\tablenotetext{a}{\citet{sol02}.}
\tablenotetext{b}{\citet{irw12a}.}
\tablenotetext{c}{Heliocentric velocity \citep{spr05}. Note that \\
this (and subsequent HI related parameters)
are \\
adjustments from \citet{irw12a} given the newer\\
 HI reference. }
\tablenotetext{d}{From the double-horned profile of \citet{spr05},\\
 corrected for redshift stretch, instrumental
effects and \\
smoothing. }
\tablenotetext{e}{\citet{spr05}, as corrected for pointing offsets, \\
source extent, and HI self-absorption. }
\tablenotetext{f}{Using $\Delta\,V$ and the inclination, above, 
assuming a spherical\\
 mass distribution, and an HI radius equivalent to the \\
optical
blue diameter of 4.8 arcmin (the HI angular extent\\
 is unknown). }
\tablenotetext{g}{CO integrated intensity using Five Colleges Radio \\
Astronomy 
Observatory data \citep{kom08}}
\end{center}
\end{table}

The radio spectrum peaks at Gigahertz frequencies, like the well-known Gigahertz-peaked spectrum (GPS) sources
seen at high redshift \citep[e.g.][]{tor01,ode98}.  Characteristic of such sources is their small radio sizes,
generally thought to be because they are young radio sources, or confined by a surrounding medium, or both 
\citep{fan09,min13}. 

{
\begin{deluxetable}{lcccccccc}
\hspace*{-10cm}
\tabletypesize{\scriptsize}
\renewcommand{\arraystretch}{0.9}
\tablecaption{Image parameters\label{table:image_parameters}}
\tablewidth{0pt}
\tablehead{
\colhead{Observation}  & \colhead{Date\tablenotemark{a}} & \colhead{$\nu_{0}$\tablenotemark{b}}
& \colhead{Beam parameters\tablenotemark{c}}& \colhead{Pixel size} & \colhead{SC Iterations\tablenotemark{d}}
&\multicolumn{3}{c}{rms\tablenotemark{e}} \\
 & &       &                                       & &             & I & Q,U & V\\
 & & (GHz) &  $\prime\prime$, $\prime\prime$, deg. &$\prime\prime$ & &\multicolumn{3}{c}{($\mu$Jy beam$^{-1}$)}\\ 
}
\startdata
{\bf CHANG-ES}  \\
\tableline
D array L-band & 30-Dec-2011  & 1.57470& 38.58, 34.27, -5.22  & 5.0 & 2A\&P & 40 & 27 & 28  \\
C array L-band  & 30-Mar-2012& 1.57484 &  12.18, 11.10, -41.76  & 1.0 & 2A\&P & 45 & 19 & 26\\
B array L-band & 11-Jun-2012 & 1.57499  & 3.51, 3.33, 22.69  & 0.5 & 2A\&P & 18 & 15 & 15 \\
\tableline
D array C-band & 19-Dec-2011 & 5.99833  & 10.98, 9.06, -1.40 & 1.0 & 6P$+$1A\&P & 15 & 15 & 12\\
C array C-band\tablenotemark{f} &23\&25-Feb-2012& 5.99854  &3.05, 2.75, -11.71 & 0.5 & 1P$+$5A\&P & 3.9 & 3.2 & 3.3 \\
\tableline
\tableline
{\bf Previous values} \\
\tableline
D array Lband (NVSS)\tablenotemark{g} & 27-Feb-1995 & 1.4 & 45, 45, 0 & 15.0 & & 450\\
Arecibo\tablenotemark{h}  & 28-Aug-1975 $\rightarrow$ 11-Sep-1975 & 2.380 & $\sim$ 162\tablenotemark{i}  
&  &   \\ 
 &  21-Dec-1975 $\rightarrow$ 03-Jan-1976 & &  \\
B array Lband (FIRST)\tablenotemark{j} & 09-Oct-1998 & 1.4351 & 6.4, 5.4, 0.0 & 1.8 & & 150\\
C array X-band\tablenotemark{k} & 21\&25Aug-1997  & 8.4 & 7.08, 5.69, 64.51 & & & 100 \\
Molonglo\tablenotemark{l} & Feb $\rightarrow$ Aug 1978 & 0.408 & 172, 172, 0 \\
\enddata
\tablecomments{New CHANG-ES data are separated from prevously measured values (where available) by the double-line.}
\tablenotetext{a}{Date (or range) that the observations were carried out (UTC, if time specified).}
\tablenotetext{b}{Central frequency of each image. Slight differences for the same band result 
from different amounts of flagging in different data sets. The bandwidths (GHz) were 1.247 $\rightarrow$ 1.503 
and 1.647 $\rightarrow$ 1.903 at L and C-bands, respectively.
}
\tablenotetext{c}{Major \& minor axis diameters and position angle of the synthesized beam.}
\tablenotetext{d}{Each application of a self-calibration iteration acted on the non-self-calibrated data, but using
a model that improved with each iteration.}
\tablenotetext{e}{rms map noise measured in a consistent fashion for all maps from random regions that do
not appear to contain background sources.  For D/L and C/L total intensity images only, the rms noise is higher
by a factor of 1.5 to 2 close to the source itself because of residual sidelobes and the presence of many background sources.} 
\tablenotetext{f}{These observations were split over two dates as specified.}
\tablenotetext{g}{Downloaded data from the NRAO VLA Sky Survey (NVSS) \citep{con98} website 
(http://www.cv.nrao.edu/nvss/)}
\tablenotetext{h}{Dressel \& Condon (1978). The observations were carried out during one of the two date intervals
specified.}
\tablenotetext{i}{Assuming a zenith angle of 10 degrees.}
\tablenotetext{j}{Downloaded data from the Faint Images of the Radio Sky at Twenty-Centimeters (FIRST)
 survey \citep{bec95} website (http://sundog.stsci.edu/index.html).}
\tablenotetext{k}{\citet{fil00}, using beam parameters corresponding to their most reliable flux density. 
Calibration errors are quoted as a few percent.}
\tablenotetext{l}{\citet{har82}.}
\end{deluxetable}}

NGC~4845 is located in the `Virgo Southern Extension'
\citep{tul82,kar13} (D = 17 Mpc, Table~\ref{table:galaxy_parameters}), providing
a unique opportunity to study
the radio/X-ray connection for this nearby AGN. 
A similar relevant case \citep{zau11}
is the radio/X-ray analysis discussing the 'birth of a relativistic
outflow' in a $\gamma$-ray transient at a redshift of $z\,=\,0.354$. NGC~4845 is
100 times closer and, with future monitoring, may provide an unprecedented opportunity
to follow the development of an outflow in what could be a 
present day lower luminosity version of a newly forming quasar.

In this paper, we present our observations and
develop a simple cone/jet model that is consistent with the data, including
the variability of the radio flux density and spectral index, the behaviour of the radio
spectrum,
the observed circular polarization, and the absence of significant linear polarization.
We also suggest a possible link between the {\it peak} X-ray and {\it peak} radio emission in the context of our 
model.

Table~\ref{table:image_parameters} provides imaging parameters from our
five data sets as well as a summary of previous radio observations from the literature, 
clearly illustrating the variability of the source.
It is not surprising that NGC~4845 has been variously listed as having
an HII/LINER type nucleus 
\citep[provided by the NASA extragalactic database (NED) based on optical spectra of][]{ho95},
a pure HII region type nucleus \citep{fil00}, and a Seyfert 2 nucleus \citep{ver10}, given
that the relative importance of the nucleus/disk changes with time.  In that respect, NGC~4845
appears to be a closer, fainter version of the recently identified `changing-look quasar' in which
the nuclear activity has changed from Type 1 to a Type 1.9 AGN over a 10 year time span \citep{lam15}.

Sect.~\ref{sec:obs_data} presents the observations and data reductions, Sect.~\ref{sec:total_intensity}
 outlines the total intensity results, in-band spectral indices, and variability of the source with fitted
spectra, and
Sect.~\ref{sec:polarization} gives the polarization results.  Sect.~\ref{sec:discussion} discusses the
core. We provide a detailed analysis of the core spectra and their variability
and also present several appendices (Appendices~\ref{appendixA}, \ref{appendixB}, and \ref{appendixC})
which, not only form a description and are consistent with the observations, but also summarize
more general concepts that may be useful for future observations of NGC4845  and other such sources, especially
in an era in which the sensitivity, fidelity  and resolution  of radio data are rapidly improving.  
 The
conclusions are presented in Sect.~\ref{sec:conclusions}.

\section{Observations and Data Reductions}
\label{sec:obs_data}

As part of the CHANG-ES program, five 
observations
were carried out (B, C, and D configurations at L-band and
C and D configurations at C-band) in all polarization products over approximately a six month period
(Table \ref{table:image_parameters}). We will refer to each observation according to
`configuration/band' (i.e. B/L, C/L, D/L, C/C, and D/C).
Data reduction and imaging details are described in \citet{irw13} and \citet{wie15}.  Here we provide a
brief
outline as well as details specific to NGC~4845.  The Common Astronomy Software Applications (CASA)
 package \citep{mcm07} was used throughout\footnote{see also http://casa.nrao.edu}.

For each observation, the galaxy was observed within
a scheduling block (SB) that included other galaxies so that the uv coverage could be maximized for each. 
 For C/C, however, the galaxy was observed
in two separate SBs on different dates which were calibrated separately and then
combined for imaging.

The calibrators were 3C~286 (the `primary' calibrator), also used for the bandpass and polarization
angle calibration, and J1246-0730 (or J1224+0330 for C/C) (the `secondary calibrator) which is near the source on the sky.
To determine the leakage terms between the R and L circularly polarized
feeds for polarization imaging, we used the unpolarized source, 
J1407+2827 = OQ~208, except for C-band for which we used the secondary calibrators which spanned a
parallactic angle $>$ 60$^\circ$ over the observations\footnote{The secondary calibrator was used in the event that
OQ~208 carried some polarization at C-band. A test using both methods, however, showed negligible differences
between the results.}.


At L-band, the frequency coverage was (in GHz) 1.247 $\rightarrow$ 1.503 and 1.647 $\rightarrow$ 1.903
(500 MHz total),
 and at C-band, it was 4.979 $\rightarrow$ 7.021 (2 GHz). 
The frequency gap at L-band was set to
avoid very strong persistent interference in that frequency range.
   
Flagging was carried out iteratively, i.e. after each round of calibration, the data were re-inspected,
more flagging was carried out if needed, the calibration tables remade,
the data then recalibrated, etc. Any flagging that was carried out for the total intensity
data (RR and LL circular polarizations) was also automatically applied to the cross-hands
(RL and LR). In addition, for the polarization calibration (Q and U maps, see below), RL and
LR data were inspected, and flagging/calibration iterations were again carried out prior to
polarization imaging.



Wide-field imaging \citep{cor08b} using the ms-mfs (multi-scale, multi-frequency-synthesis) algorithm 
\citep{cor08,rau11} was carried out, with simultaneous fitting of a simple power law 
($S_\nu\,\propto\,\nu^\alpha$) across the band. 
Since NGC~4845 is point-like (Sect.~\ref{sec:total_intensity}), however, the use of many scales was not warranted.
After a variety of trials, we found that a classic clean gave the best results for highly
unresolved cases (D/L, C/L, D/C) with 
the addition of modest scales for slightly resolved cases (B/L: 0, 5, 10, and 20 arcsec; 
C/C: 0 and 5 arcsec). 

For all maps, Briggs robust = 0 uv weighting was used \citep{bri95},
as implemented in CASA. 
A variety of self-calibration steps were then employed, either phase-only (P) or amplitude
and phase together (A\&P) depending on the specific data set and what was required to
improve
the dynamic range of the map.
Care was taken to ensure that, with successive iterations,
 the self-calibration did not go so deep as to start including
spurious values such as negatives or residual sidelobes and the peak specific intensity did not
decline.  
The primary beam (PB) correction was then applied. 
Table \ref{table:image_parameters} summarizes the imaging parameters.

 For each observation, imaging of Stokes Q, U, and V was also carried out
for each observation using the same self-calibration tables and the same
 input parameters as for total intensity  (except B/L for which the scales were
reduced to
0, 5, and 20 arcsec). 
Maps of linear polarization, $P\,=\,\sqrt{Q^2\,+\,U^2}$ were then made, corrected for the bias introduced by the
fact that P images do not obey Gaussian statistics \citep[e.g.][]{sim85,vai06}
as well as maps of polarization angle.
 P and V maps were then corrected for the primary beam.

In the case of circular polarization,
 we also remade the maps without fitting
a spectrum across the band, essentially averaging all spectral channels together without attempting to fit a spectral
index.
However, those maps were significantly worse 
than when a spectral fit was carried out.  For example, 
at C/L, including a spectral fit increased the peak specific intensity by 5\% and decreased the rms noise by 32\%. We
therefore
 retained the $V$ maps that included a spectral fit.

For each observation,
{\it in-band} spectral index, $\alpha$, maps and 
$\alpha$ {\it error} maps were also made for each image \citep{rau11}, as implemented in
the CASA {\it clean} algorithm.  Following this, the post-imaging 
prescription of \citet{wie15} was followed
 which ensures that the
images are corrected for 
the spectral index that is imposed by the primary beam \citep[e.g. see also][]{irw12b,irw13}
and smooths over artifacts.
A departure from that prescription, however, is that we have been more conservative
in our cut off, i.e.
spectral index maps for NGC~4845 have been cut off wherever the total intensity
falls below 10$\sigma$ instead of 5$\sigma$ (the latter is the norm for CHANG-ES).

A 
weighted mean spectral index, $\overline{\alpha}$, and its associated error, $\overline{\Delta\alpha}$,
were calculated over a region of interest, taken to be the  {\it FWHM of a single beam only}
so, essentially, we are adopting a measure of
$\alpha$ in a single beam at the galaxy center, a region in which variations in the spectral index have been minimized.
The values are calculated via,

\begin{equation}
\overline{\alpha}\,=\,\left\{\frac{{\Huge{\Sigma_i}}\,\left(\frac{\alpha_i}{{\Delta\alpha_i}^2}\right)}{{\Huge{\Sigma_i}}\,\left(\frac{1}{{\Delta\alpha_i}^2}\right)}\right\}
\end{equation}

\begin{equation}
\overline{\Delta\alpha}\,=\,\left\{\frac{N_{p/b}}{{{\Huge{\Sigma_i}}\,\left(\frac{1}{{\Delta\alpha_i}^2}\right)}}\right\}^{1/2}
\end{equation}

\noindent where the summations are over the map pixels in the region of interest (i.e. the central beam).  
The factor, $N_{p/b}$ is the number of pixels per beam which increases the error since the pixels are not independent.
These weightings are adopted to ensure that pixels with the highest S/N and lowest uncertainty are given the highest weight.
Since the source is mainly unresolved (Sect.~\ref{sec:total_intensity}), the weighting is close to that of a Gaussian.
The beam size is shown at lower left in Figs.~\ref{fig:spectral_indices}
and \ref{fig:spectral_indices_C}. 

The results are given in
Table~\ref{table:spectral_indices}. 
It is important to point out that the dynamic range (DR) of each total intensity map is  very high.
The DRs are 5600/1 (D/L), 5460/1 (C/L)
 and 11,600/1 (B/L) at L-band
(but note that for D/L and C/L, the rms is higher close to the source,
 see notes to Table~\ref{table:image_parameters}). At C-band, the DR is higher still, namely
28,300/1 (D/C) and 88,900/1 (C/C).  As a result, the formal errors are quite small.  However,
tests have shown that the true error in the spectral index for a {\it typical} CHANG-ES
galaxy is $\approx\,$20\% higher \citep{wie15}.  Also, for our C/L and D/L data sets, if we shift the measuring region
up or down slightly from the center, then $\overline{\alpha}$ will change because of the north-south
gradient observed for those two data sets (Fig.~\ref{fig:spectral_indices}). We have ensured that all data
sets are measured at the same position and in the same way and in Sect.~\ref{sec:spectra} we 
outline a number of checks indicating that possible variations in alpha do not affect the conclusions of this paper.

Spectral index maps for the circularly polarized emission (not shown) were formed and measured in the same way.

\begin{deluxetable}{lcccc}
\tabletypesize{\scriptsize}
\tablecaption{In-band Spectral Indices\label{table:spectral_indices}}
\tablewidth{0pt}
\tablehead{
\colhead{Observation} & \colhead{Date}  &
\colhead{$\overline{\alpha}$} &\colhead{$\overline{\alpha_{AGN}}$} & \colhead{$\overline{\alpha_V}$}  \\  
}
\startdata
D array L-band & 30-Dec-2011 (T1) & +0.95 $\pm$ 0.01 & +1.11  $\pm$ 0.02 & -2.2 $\pm$ 0.1\\ 
C array L-band & 30-Mar-2012 & +0.97 $\pm$ 0.01 & +1.11  $\pm$ 0.02 &-2.5 $\pm$ 0.2\\
B array L-band & 11-Jun-2012 &  +0.81 $\pm$ 0.17 &  +0.81 $\pm$ 0.17 & -3.4 $\pm$ 0.1\\
\tableline
D array C-band & 19-Dec-2011 & -0.453 $\pm$ 0.003 & -0.448 $\pm$ 0.006&\\
C array C-band & 23\&25-Feb-2012 &  -0.493 $\pm$ 0.001 &-0.493 $\pm$ 0.001 &\\
\enddata
\tablecomments{Weighted mean spectral indices of the total intensity emission, $\overline{\alpha}$, 
the unresolved core only, $\overline{\alpha_{AGN}}$,
and the circular polarization, $\overline{\alpha_V}$.  Measurements were made only for the region covered by the
half-power beam width at the galaxy center.
For quoted error for $\overline{\alpha}$
is the formal random error only; typical errors are likely $\approx$ 20\% higher (see Sect.~\ref{sec:obs_data}).
The quoted error for $\overline{\alpha_{AGN}}$ has been doubled from the previous column for resolutions which required
an adjustment from  $\overline{\alpha}$ to $\overline{\alpha_{AGN}}$
(Sect.~\ref{sec:spectral_indices}). For $\overline{\alpha_V}$, the uncertainties may be
much higher than the formal error and thus all values of $\overline{\alpha_V}$ are taken to be roughly the same.
}
\end{deluxetable}


\section{Results}
\label{sec:results}

\subsection{Total Intensity Emission}
\label{sec:total_intensity}

\begin{figure*}[!ht]
   \includegraphics*[width=1.0\textwidth]{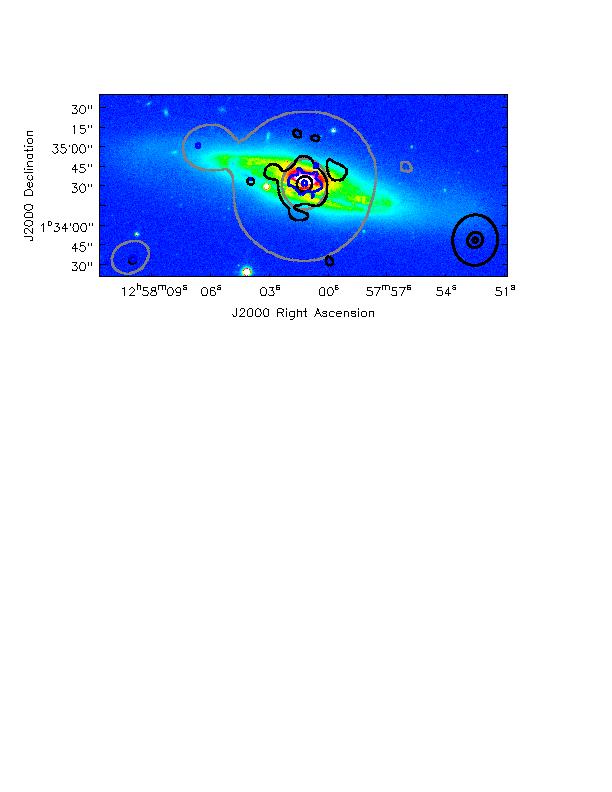}
   \hspace{-1.20in}
    \vspace{-4truein}
   \caption{L-band contours on a Sloan Digital Sky Survey (dr7) r-band image of NGC4845. 
Contours are set at 10$\sigma$ and 50\% of the peak for each data set.  D/L: grey (0.40, 112.4 mJy beam$^{-1}$),
C/L: black (0.45, 122.8 mJy beam$^{-1}$), and B/L: blue (0.18, 104.7 mJy beam$^{-1}$).  The beam sizes are shown
at lower right.
}
\label{fig:optical_Lband_contours}
\end{figure*}

In Fig.~\ref{fig:optical_Lband_contours}, we show simple contours (10$\sigma$ and half-maximum only) 
of the L-band emission for all 3 array
configurations on  
an r-band image of NGC~4845 from the Sloan Digital Sky Survey (SDSS)\footnote{http://www.sdss.org/}.
It is clear that the radio emission is unresolved (cf. the beams at lower right)
with the exception of a small disk around the nucleus showing up at the highest resolution
(B/L).  We will refer to this disk as the `central disk'.
In no case, however, do we see emission from
an extended disk (Sect.~\ref{sec:no-disk}), i.e. over a size that might approach the optical extent of the galaxy. 

\begin{figure*}[!ht]
   \centering
   \includegraphics*[width=0.5\textwidth]{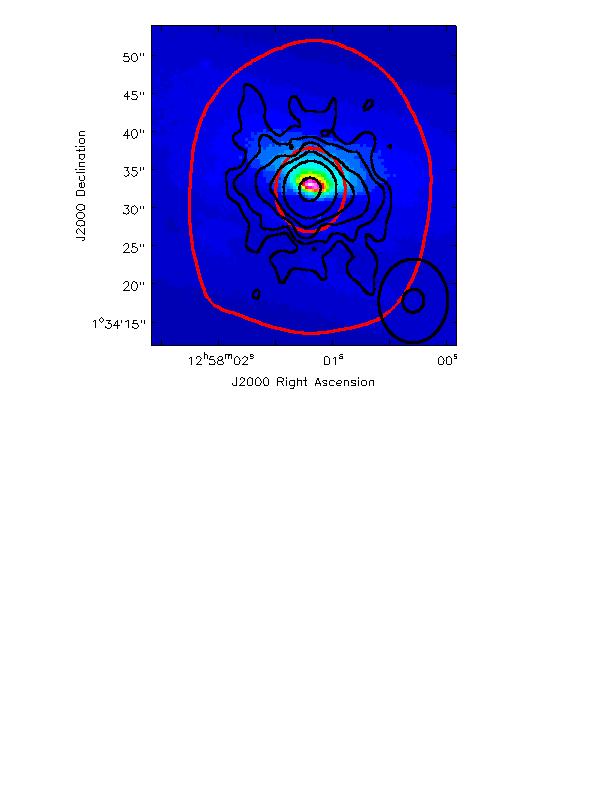}
   \hspace{-1.20in}
    \vspace{-2truein}
   \caption{C-band contours on a blow-up of the optical emission
as shown in Fig.~\ref{fig:optical_Cband_contours}. 
Contours are red for D/C at 0.15 (10$\sigma$) and 212.4 mJy beam$^{-1}$ (half-peak value).  Contours are
black for C/C at 0.039, 0.10, 0.30, 1.00, 6.00, and 173.1 mJy beam$^{-1}$ (half-peak value).  The beam sizes are shown
at lower right.
}
\label{fig:optical_Cband_contours}
\end{figure*}

Fig.~\ref{fig:optical_Cband_contours} shows C-band contours on a blow-up of the same optical emission.
 This image clearly shows asymmetric optical emission in the shape of
a cone to the north of the nucleus. The galaxy is slightly inclined (Table~\ref{table:galaxy_parameters}) such that
its southern edge is closest to us; consequently, a corresponding southern optical cone, if it exists, would
be obscured by the disk.

Again, the D/C (red contours) radio emission is unresolved but the C/C emission (black contours)
delineates the small central disk quite well. 
 The disk subtends 21.7 arcsec (1.8 kpc in diameter)
to the displayed 10$\sigma$ contour level at a position angle of 83$^\circ$, in contrast to the larger-scale
optical disk which shows some distortion and whose position angle varies from 74 to 80 degrees, depending on
the isophotes to which it is measured. 
A similarly angled radio disk is visible at 8.4 GHz, whose size ($\approx$
20 arcsec) agrees with our result \citep{fil00}. 

The central disk shows some extensions to the north and south (B/L and C/C,
Figs.~\ref{fig:optical_Lband_contours} and \ref{fig:optical_Cband_contours}) which could be related to
outflow.  \citet{fil00} similarly note a small northern extension at 8.4 GHz.

\begin{figure*}[!ht]
   \centering
   \hspace{-0.2in}
   \includegraphics*[width=0.4\textwidth]{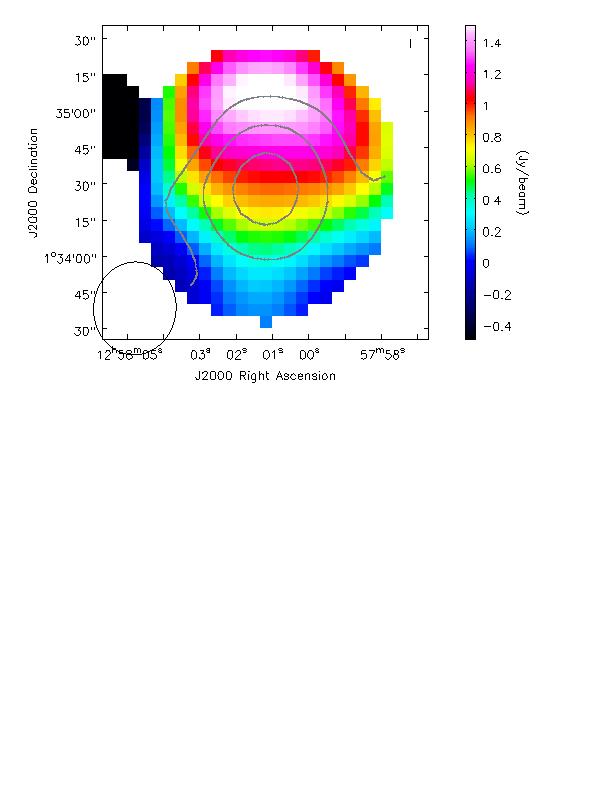}
   \hspace{-0.3in}
   \includegraphics*[width=2.5in]{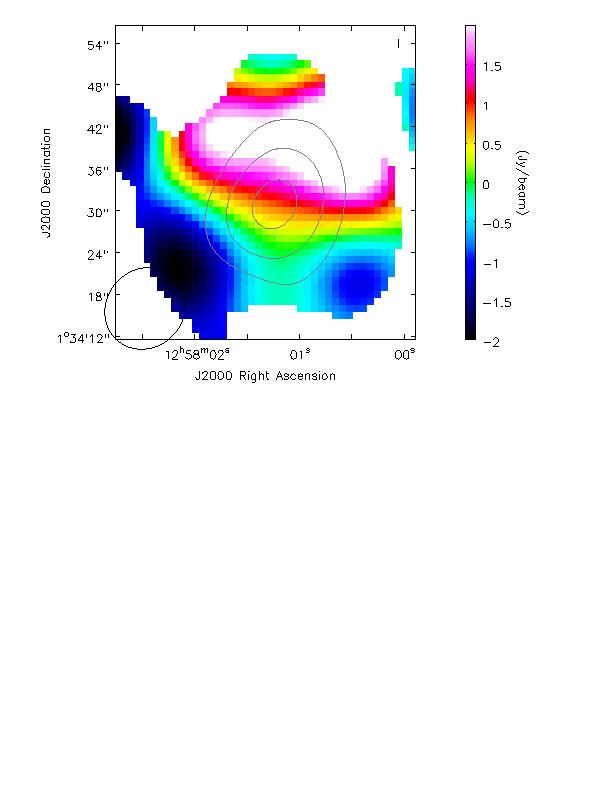}
   \hspace{-0.2in}
   \vspace{-1.5truein}
   \begin{minipage}{0.2\textwidth} 
   \vspace{-3.5truein}
   \includegraphics*[width=1.75in]{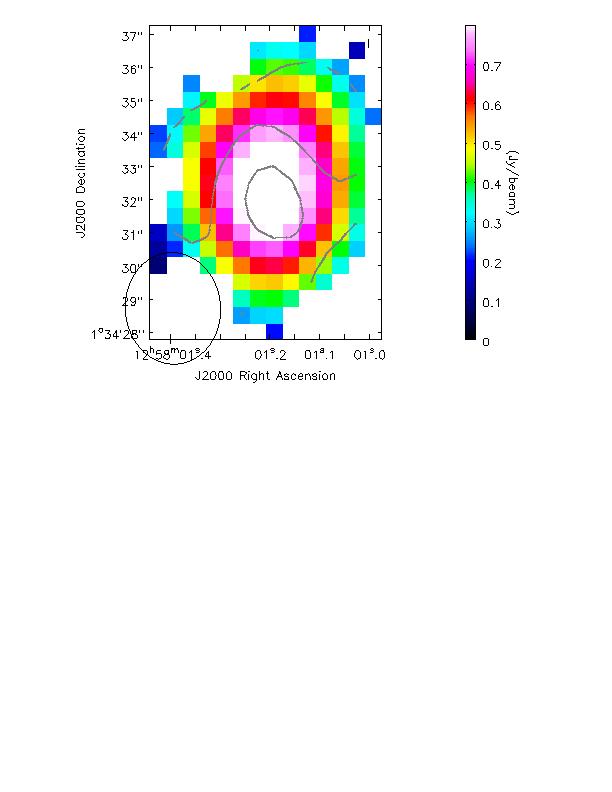}
   \end{minipage}
   \caption{L-band spectral index maps formed as described in Sect.~\ref{sec:obs_data}.  Colours represent the spectral
index with the colour legend shown at right.  Grey solid contours represent values from the corresponding 
error map, with smallest values in the interior to larger values towards the outside.  {\it Note that the size and
colour scales are different from map to map.}  The ellipse at lower left shows the beam.
{\bf Left:} D/L spectral index map with error contours at 0.01, 0.025, and 0.05. 
{\bf Center:} C/L spectral index with error contours at  0.01, 0.025, and 0.05.
{\bf Right:} B/L with error contours at 0.15 and 0.2 (south-west contour is 0.15 and broken north contour is 0.2.}
\label{fig:spectral_indices}
\end{figure*}

\begin{figure*}[!ht]
   \centering
   \includegraphics*[width=0.4\textwidth]{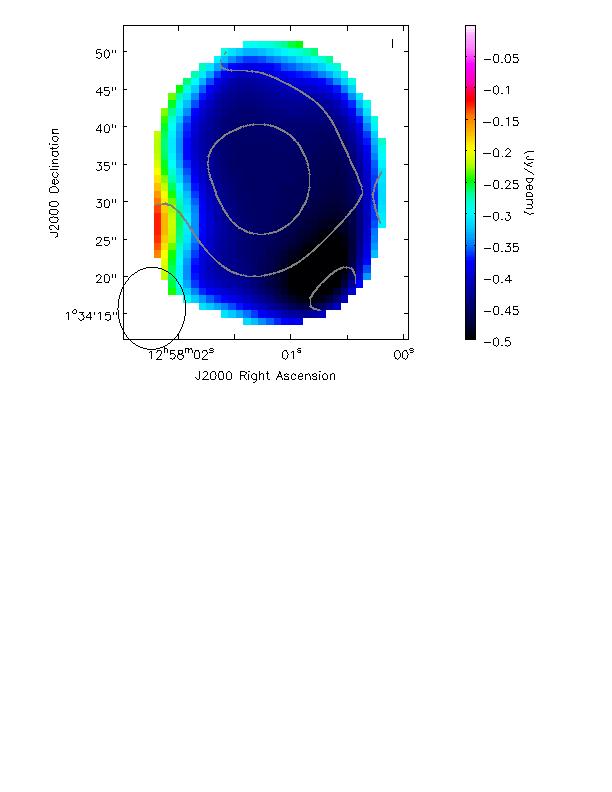}
   \begin{minipage}{0.3\textwidth}
   \vspace{-3.5truein}
   \includegraphics*[width=2.2in]{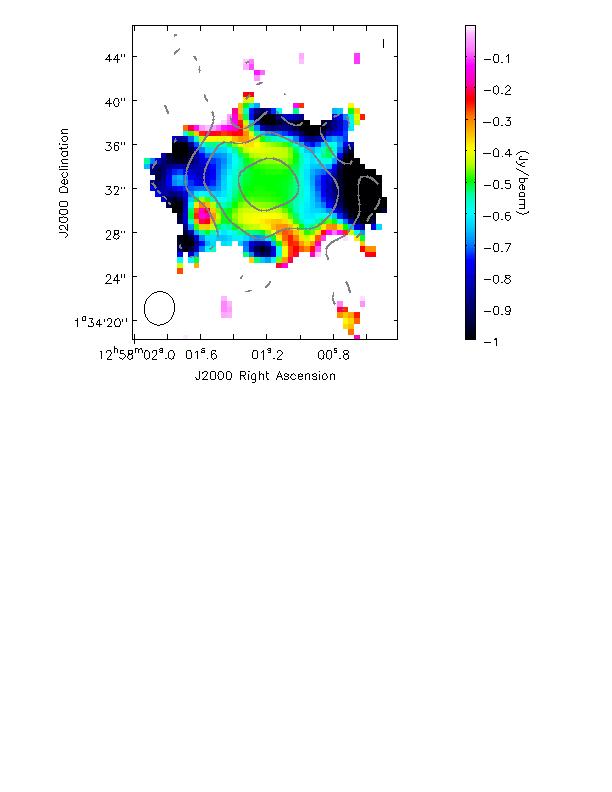}
   \end{minipage}
   \hspace{-0.20in}
   \vspace{-1.5truein}
   \caption{C-band spectral index maps formed as described in Sect.~\ref{sec:obs_data}.  Colours represent the spectral
index with the colour legend shown at right.  Grey solid contours represent values from the corresponding 
error map, with smallest values in the interior to larger values towards the outside.  {\it Note that the size and
colour scales are different from map to map.}  The ellipse at lower left shows the beam.
{\bf Left:} D/C spectral index map with error contours at 0.01, 0.05, and 0.1. 
{\bf Right:} C/C with error contours at 0.01, 0.1, and 0.2.} 
\label{fig:spectral_indices_C}
\end{figure*}

\subsubsection{Flux Densities} \label{sec:flux_densities}

Flux densities were measured from the primary beam corrected images
and are given in Table~\ref{table:flux_densities} along with previous literature values.
With the exception of the 
D/C data set (see below), the quoted errors are dominated by variations in flux density that result from
adjusting the box size around the source. 
It is abundantly clear that NGC~4845 is a variable
radio source, its L-band flux density, $S$ having increased approximately 6-fold since the NVSS survey
results in 1995.

\begin{deluxetable}{lcccccccc}
\tabletypesize{\scriptsize}
\tablecaption{Flux densities\label{table:flux_densities}}
\tablewidth{0pt}
\tablehead{
\colhead{Observation} & Frequency & \colhead{Time\tablenotemark{a}} &\colhead{$S$\tablenotemark{b}} & 
\colhead{$S_{AGN}$\tablenotemark{c}} &
\colhead{$P$\tablenotemark{d}} & \colhead{$V$\tablenotemark{e}} 
&\colhead{$P/S$\tablenotemark{f}} & \colhead{$V/S$\tablenotemark{f}}\\
 & (GHz) & (days) & (mJy) & (mJy) & (mJy) & (mJy)  & \% & \%
}
\startdata
D array L-band & 1.57 & 0 &230 $\pm$ 2 & 211 $\pm$ 3 & 0.55 $\pm$ 0.05 & 6.6 $\pm$ 0.2 & 0.24 $\pm$ 0.02 & 2.9 $\pm$ 0.1\\
C array L-band & 1.57 & 91 &260 $\pm$ 2 & 241 $\pm$ 3 & 0.35 $\pm$ 0.03 & 5.7 $\pm$ 0.1 & 0.13 $\pm$ 0.01 & 2.2 $\pm$ 0.06\\
B array L-band & 1.57 & 164 &238 $\pm$ 1 & 219 $\pm$ 4 & 1.2 $\pm$ 0.5 & 5.5 $\pm$ 0.4 & 0.5 $\pm$ 0.2 & 2.3 $\pm$ 0.2\\
\tableline
D array C-band & 6.00 & -11 & 432 $\pm$ 2 & 425 $\pm$ 3 & 2.3 $\pm$ 0.9 & $<$ 0.036 & 0.5 $\pm$ 0.2 &  \\
C array C-band & 6.00 & 56 & 362 $\pm$ 1 & 355 $\pm$ 2 &0.4 $\pm$ 0.2 & $<$ 0.010 & 0.11 $\pm$ 0.06 & \\
\tableline
\tableline
D array L-band (NVSS)\tablenotemark{g} & 1.4 & &46.0 $\pm$ 1 &  \\
B array L-band (FIRST)\tablenotemark{g} & 1.4 &  & 33.9 $\pm$ 0.4 & \\
Arecibo\tablenotemark{h} & 2.38 & & 31 $\pm$ 3\\
C array X-band\tablenotemark{h}  &  8.4 & & 12.5\\
Molonglo\tablenotemark{h} & 0.408 & & 193 $\pm$ 19 & &\\
\enddata
\tablecomments{Flux densities for the data sets of Table~\ref{table:image_parameters}. Frequencies have been 
rounded.
Upper limits assume that the source is unresolved and that a detection would require
a 3$\sigma$ signal, where $\sigma$ is the rms noise of the map (Table~\ref{table:image_parameters}).}
\tablenotetext{a}{Time in days since T1 = 30-Dec-2011.}
\tablenotetext{b}{Flux density of the total intensity emission.  The quoted 
error does not include the 1\% absolute flux density error of 3C~286. See Sect.~\ref{sec:flux_densities} for details.}
\tablenotetext{c}{Flux density of the AGN after subtracting off the disk emission. The latter was determined by
subtracting off Gaussian
fits to the peak of the highest resolution data at each frequency (Sect.~\ref{sec:flux_densities}).}
\tablenotetext{d}{$\,$Linearly polarized flux density, $P\,=\,\sqrt{Q^2+U^2}$,
where P has been corrected for positive bias.  $P$ has been measured in varying-sized regions
and, where the emission was point-like (D/L and C/L), via Gaussian fits as well. The quoted error reflects
the resulting variations. 
}
\tablenotetext{e}{Flux density of the circularly polarized emission. Errors are random only, i.e. as evaluated
from the rms noise and residual sidelobes in the map itself. See Sect.~\ref{sec:polarization}. }
\tablenotetext{f}{Percentage polarization of either the linearly or circularly polarized emission.  For $V/S$, the
C-band results have been omitted since the $V$ values are non-detections.}
\tablenotetext{g}{Flux densities from downloaded data sets (FIRST and NVSS) were measured in the same way as for our data. 
The NVSS value agrees
with the result from Condon et al. (1998) and the FIRST value agrees with \citet{bec12}. }
\tablenotetext{h}{See Table~\ref{table:image_parameters} for references.}
\end{deluxetable}

Since we will also argue that the flux density has varied over the several-month
 timescale of our own observations,
it is important to consider possible uncertainties that might not have been taken into account
from the map measurements.
These include a) missing
flux due to missing spatial scales, b) calibration errors, 
 and c)
changes in flux density from self-calibration. 

For a), the largest angular scale detected at L-band is 16 arcmin and, at C-band, is 4 arcmin.  The
structures shown in 
Figs.~\ref{fig:optical_Lband_contours} and \ref{fig:optical_Cband_contours} are
well within these limits, indicating that all flux has been detected in each configuration/frequency-band
combination. Calibration errors (b), as measured during the transfer from the primary to the secondary
calibrator, are less than the errors quoted in Table~\ref{table:flux_densities} except for D/C for which
they dominate and are quoted in the table. 
The uncertainties of
Table~\ref{table:flux_densities} do not take into account uncertainties in the primary calibrator 
(3C~286) itself \citep[of order one percent]{per13},
but this will not affect comparisons between data sets. 
Self-calibration (c) can result in some modification of the flux density. 
The largest such change occurred for the C/L data set, for which a 4 mJy difference is observed between the
non-self-calibated data (256 mJy) and the self-calibrated data (260 mJy), an increase of
1.5\%. Only a decrease in flux density
would be a concern and any decrease observed had a smaller relative error than this. 
Self-calibration, for all data sets, resulted in an increase in dynamic range and a lower rms
noise. 

Since the emission was point-like, the
flux densities were also measured by fitting a Gaussian to the emission; the result agreed
with the tabulated values, except for the high resolution data (B/L and C/C) for which
some disk-emission is also seen and therefore the Gaussian fitted flux densities slightly underestimate the total flux
density (92\% and  98\%, for B/L and C/C, respectively.  
This also allows us to separate the flux density of the 
non-varying disk from the 
AGN, thereby isolating the flux density of the variable AGN alone, $S_{AGN}$
(Table~\ref{table:flux_densities}).  From the global flux densities of the disk alone at the two frequencies
(19 $\pm$ 4 mJy at L-band and  7 $\pm$ 1 mJy at C-band) we find that the
spectral index for the disk is $\alpha_{disk}\,=\,-0.74$.  Such a value is typical of galaxy disks, in general.

\subsubsection{Core Size and Astrometry}\label{sec:astrometry}

\begin{table*}[ht]
\scriptsize
\begin{center}
\caption{Source astrometry\label{table:astrometry}}
\begin{tabular}{cccc}
\tableline\tableline
{Observation} & {$\nu_0$\tablenotemark{a}} 
& {RA, DEC (J2000)\tablenotemark{b}}
& {Source size\tablenotemark{c}} \\
 & (GHz) & h:m:s, $\circ$:$\prime$:$\prime\prime$
&  $\prime\prime$, $\prime\prime$, deg.\\
\tableline
B array L-band & 1.57499 & 12:58:01.1955 $\pm$ 0.0007, +01:34:32.42 $\pm$ 0.01
& 0.67 $\pm$ 0.03, 0.15 $\pm$ 0.09, 87 $\pm$ 170\\
\tableline
C array C-band & 5.99854 &
~12:58:01.1981 $\pm$ 0.0001, +01:34:32.424 $\pm$ 0.001 & 0.40 $\pm$ 0.03, 0.36 $\pm$ 0.03, 97 $\pm$ 175 \\
\tableline
\end{tabular}
\tablenotetext{a}{Reference frequency for each image, as in Table~\ref{table:image_parameters}.}
\tablenotetext{b}{Positions from Gaussian fits. The quoted uncertainty is the larger 
of the formal errors or variations resulting from the adoption of
three different boxes within which the fit was carried out.}
\tablenotetext{c}{Deconvolved from the beam, from Gaussian fits. The quoted uncertainty is as described in
note {\it b}. The fit includes an offset to account for the fact that the AGN is within a disk.}
\end{center}
\end{table*}

The Gaussian fits discussed in the previous section were also used to determine the size and
astrometry of the central core (which we take to represent the AGN),
using the highest resolution data (C/C and B/L,
Table~\ref{table:astrometry}). 
 The average central position of the source is at
 RA = 12$^{\rm h}$58$^{\rm m}$01.1968$^{\rm s}$ $\pm$ 0.0007$^{\rm s}$ ($\pm\,0.01^{\prime\prime}$),
DEC = 01$^\circ$34$^\prime$32.42$^{\prime\prime}$ $\pm$ 0.01$^{\prime\prime}$, where we adopt the higher
of the uncertainties\footnote{These are formal errors from the CASA {\it imfit}
task.} from the two data sets.  This is the most precise position yet determined
for the core of NGC~4845 and should also be the most accurate, given that radio data do not
suffer from extinction as do optical data.  Nevertheless, our position agrees well with that of
the SDSS (uncertainty of $0.2^{\prime\prime}$) as quoted in NED.

The size of the central core, after deconvolving from the synthesized beam,
is also provided.
 A comparison with the beam sizes given in  Table~\ref{table:image_parameters}
shows that {\it the central core is unresolved, as measured in the highest resolution
data sets}.  (Repeating this exercise for the lower resolution data gives the same conclusion.)
The deconvolved source size ($0.2\,\rightarrow 0.7$ arcsec) is also of order or smaller than
the {\it pixel} size (0.5 arcsec = 41 pc) which sets a limit on the 
possible precision of size measurable by a Gaussian fit.  Therefore, the unresolved core
could be smaller still.

\subsubsection{In-band Spectral Indices} \label{sec:spectral_indices}

Since the source is variable (Sect.~\ref{sec:variability}), a `classical' computation of
spectral index {\it between} L-band and C-band  
would not yield reliable results since the two different bands were measured at different times.  Fortunately,
the wide bands used in the CHANG-ES observations make it possible to compute {\it in-band} spectral
indices (Sect.~\ref{sec:obs_data}). 
Figs.~\ref{fig:spectral_indices} and \ref{fig:spectral_indices_C} show the resulting maps for L-band
and C-band, respectively.

Fig.~\ref{fig:spectral_indices} shows successively higher resolution (left to right) L-band spectral index
maps. A largely north-south gradient is visible in the lower resolution
D/L and C/L images, though not in the high resolution B/L data. These gradients are intriguing, since they
appear to align in the direction of the optical cone (Sect.~\ref{sec:total_intensity}) and since they are
visible in
two independent data sets.  
Since the gradients become most pronounced after self-calibration,
 a variety of tests were run with different self-calibration iterations, inputs
and box sizes; nevertheless, the gradient remains.  
To be cautious, however, we
have computed the weighted average spectral indices, $\overline\alpha$
 (see details, Sect.~\ref{sec:obs_data}), only for the central FWHM of the emission
shown in  Fig.~\ref{fig:spectral_indices}.  That is,  $\overline\alpha$ applies to a size equivalent to 
the beam size shown at lower left
 rather than to the entire region shown in the figures.  This also ensures that the spectral
indices are measured only in regions where the error (as shown by contours) is lowest  and
emission is highest.
The results are given in
Table~\ref{table:spectral_indices}. {\it Note that L-band spectral indices in all configurations are positive, and of
order +1.}




Fig.~\ref{fig:spectral_indices_C} shows 
the C-band spectral indices.
The weighted mean spectral indices, again for the central FWHM of the emission, are given in
Table~\ref{table:spectral_indices}.  {\it C-band spectral indices are all negative, of order -0.5.}

It is clear that the spectral index of this compact core 
peaks {\it between} L-band ($\nu\,=\,1.57$ GHz) and C-band
($\nu\,=\,6.00$ GHz).  The source therefore may be a nearby lower-luminosity analogue of 
the GPS sources seen at higher redshifts \citep[e.g.][]{ode91}.

The importance of measuring in-band spectral indices cannot be overstated for these observations. 
Since the source is variable, a classical measurement of spectral index {\it between} the centers of the two bands
would have given grossly erroneous results, {\it even if the observations had been simultaneous}, since
a straight line (in log space) would simply have been drawn between the flux densities of the mid-points of
the two bands (see also Fig.~\ref{fig:spectra}). Our in-band spectral indices correspond to the
{\it slopes} of the spectrum at the midpoints of both bands, a fact that we will exploit in Sect.~\ref{sec:spectra}. 


Finally,  the spectral index of the weak disk emission, $\alpha_{disk}\,=\,-0.74$ (Sect.~\ref{sec:flux_densities}) is not the same
as that of the central core in either band. 
 Since the HPBW region  in which   $\overline{\alpha}$ was measured for the low resolution data
 (D/L, D/C, and C/L) encompasses both the core and the disk, a small
correction is required to obtain the spectral index of the core itself.  
For each band, this was done by determining the flux density of the disk at the upper and lower ends of the
band using  $\alpha_{disk}$, determining the total flux density for the upper and lower ends of the band using
  $\overline{\alpha}$, subtracting off the disk flux density from the total at the two band ends, 
and recalculating the in-band spectral indices, yielding  $\overline{\alpha_{AGN}}$. These results are also given
in Table~\ref{table:spectral_indices}

\subsubsection{Variability} \label{sec:variability}

Table~\ref{table:flux_densities} reveals a significant change in flux density
with time for NGC~4845.  

For the historical data at L-band,
the flux density appears to
decrease from the 1995 NVSS (D-array) value of 46 mJy to the 1998 FIRST 
(B-array) value of 34 mJy. 
However, the FIRST B-array data 
 reveals only a point source without the disk since the rms noise is of order
the brightness of the disk as measured in the CHANG-ES B/L data set.
 The difference between those two flux densities (12 mJy) may therefore
be attributable to some missing low brightness extended emission of the central disk
which we have measured to be 19 $\pm$ 4 mJy (Sect.~\ref{sec:flux_densities}).
Two additional earlier archival VLA data sets are also available, though unpublished, which we have
now reduced.  These 
are L-band D-array data (Mar. 23, 1988) and  L-band B-array data (Nov. 15, 1987) 
for which the flux densities are 48 mJy and 28 mJy, respectively.  Given uncertainties
(especially at the earlier epoch),
we consider these values to
 agree with the NVSS and FIRST results, respectively.  Therefore, 
of the limited data available, there was no detected L-band variability prior to 1998.

On the other hand, our CHANG-ES observations have now shown
an {\it increase} at L-band of approximately a factor of 6 compared to the NVSS results.
Unfortunately, there is no prior measurement of C-band flux density
in the literature.  The (1986 to 1987) Green Bank 4.85 GHz survey
\citep[GB6,][]{gre96}, which covered this region of sky to a limit of $~\approx\,42$ mJy (their Fig. 2), 
did not detect the source.  Allowing for the $~5\sigma$ requirement for the source to be included in
their list suggests that the C-band flux density was $<\,210$ mJy at the earlier epoch, or a factor of 
at least 2 lower than
our peak C-band measurement.  We suspect that the flux density change was much higher, though, given the L-band measurements
and the fact that the source is highly absorbed at L-band but not C-band
 (Sect.~\ref{sec:lowfreqturnover}).



\begin{figure*}[!ht]
   \centering
   \includegraphics*[width=0.8\textwidth]{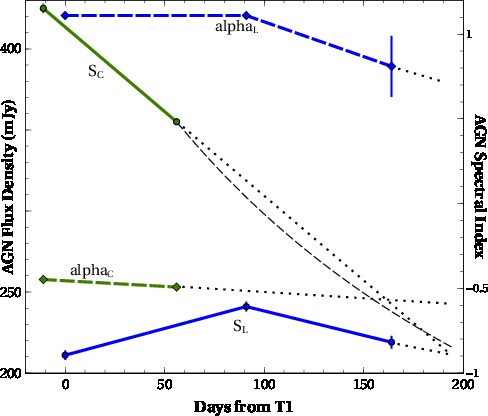}
   \hspace{-0.0in}
   \vspace{-0.1truein}
   \caption{AGN flux densities, $S_{AGN}$ (solid lines with circular markers), 
and in-band spectral indices, $\overline\alpha$ 
(dashed lines with diamond markers), are shown for the
L-band (blue) and C-band (green) data, from Tables~\ref{table:flux_densities} and 
\ref{table:spectral_indices}, as a function of time, where T1 = 30-Dec-2011. Error bars are shown but are mostly smaller than
the marker size. 
 Dotted lines show linear extrapolations of the
curves to the date, 13-Jul-2012 (T1 + 196 days), at which the L-band and C-band flux densities are equal
 (Table~\ref{table:matching_times}). The dashed curve shows an extrapolation of the C-band flux density of the form,
$S_C\,\propto\,t^{-5/3}$ (Eqn.~\ref{eqn:surprisingresult}).
} 
\label{fig:variability}
\end{figure*}

A plot of AGN flux density and 
in-band spectral indices (both corrected for the disk)
as a function of time for the CHANG-ES data is shown in Fig~\ref{fig:variability}.
Earlier historical data have been omitted, given the large lapse in time. This plot shows
that the AGN flux density has varied even over the few
month time period of our observations.  At L-band, the flux density increases, and then decreases.  At
C-band, the flux density shows a strong decline. 
The spectral indices (dashed curves) also reveal variations in both bands.

From these data, we adopt 3 time stamps, namely
Time 1 (T1): the first time at which L-band data are obtained and for which C-band data can be
interpolated from its curve, Time 2 (T2 = T1 + 56 days): 
the last time at which C-band data have been obtained and for
which L-band data can be interpolated, and Time 3 (T3 = T1 + 196 days): 
an extrapolation of both L-band and C-band data to
the time at which the flux densities at both bands are equal.  In all cases, simple linear interpolations
or extrapolations are performed, and the same is done for the spectral indices.
The resulting data  (Table~\ref{table:matching_times}) provide flux densities and
spectral indices for matching times.

\begin{deluxetable}{lcccccc}
\tabletypesize{\scriptsize}
\tablecaption{Flux Densities and Spectral Indices for Matching Times\label{table:matching_times}}
\tablewidth{0pt}
\tablehead{
\colhead{Date} & \multicolumn{3}{c}{L-band} & \multicolumn{3}{c}{C-band}\\
\tableline
\colhead{} & \colhead{$S_{AGN}$} & \colhead{Frequency} &\colhead{$\overline{\alpha_{AGN}}$}
& \colhead{$S_{AGN}$} & \colhead{Frequency} &\colhead{$\overline{\alpha_{AGN}}$}  \\  
 & \colhead{mJy} & \colhead{GHz} & & \colhead{mJy} & \colhead{GHz} &}
\startdata
T1 (30-Dec-2011)                 & 211  & 1.575 & +1.11   & 414 & 5.998 &  -0.455  \\
T2 (24-Feb-2012 = T1 + 56 days)  & 229  & 1.575 & +1.11  & 355 & 5.999 &  -0.493 \\
T3 (13-Jul-2012 = T1 + 196 days) & 209  & 1.575 & +0.68   & 209 & 5.999 &  -0.587 \\
\enddata
\tablecomments{$\,$L-band and C-band flux densities and spectral indices for matching times, interpolated or extrapolated from
Fig.~\ref{fig:variability} (see also Sect.~\ref{sec:variability}).
}
\end{deluxetable}

\subsubsection{The AGN Spectrum and its Variability} \label{sec:spectra}

From the data of Fig.~\ref{fig:variability}, it is possible, not only to estimate the spectrum of the AGN,
but also to see how it varies with time. 
For each time stamp, we now have four data points that can be used to constrain the spectrum: 
the L-band flux density, the L-band spectral index, the C-band flux density, and the
C-band spectral index, where the spectral indices represent tangent slopes of the spectrum at the two frequencies.  We then
fit a polynomial to the spectrum of the form
\begin{equation}
S_\nu\,=\,a_0\,+\,a_1\,\nu\,+\,a_2\,\nu^2\,+\,a_3\,\nu^3
\label{eqn:polynomial}
\end{equation}
where $S_\nu$ is the flux density at frequency $\nu$ and $a_i$ are constants.
The derivative of this equation is equated to the derivative of the flux density
equation, $S_\nu\,\propto\,\nu^\alpha$, both of which represent
tangent slopes to the spectrum at frequency, $\nu$.  With two flux densities and two slopes at
each time stamp, Eqn.~\ref{eqn:polynomial} was solved for the spectrum at each time stamp.

\begin{figure*}[!ht]
   \centering
   \includegraphics*[width=0.7\textwidth]{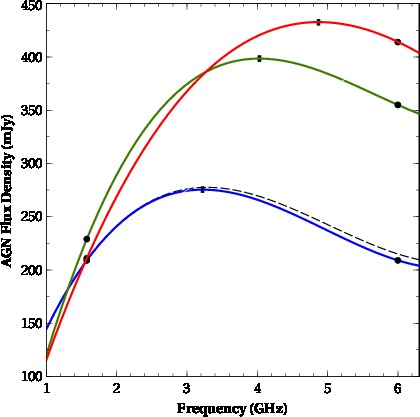}
   \caption{Spectra from polynomial fits (Table~\ref{table:spectral_fits})
 for three time stamps: T1 (30-Dec-2011) in red, T2 = T1 + 56 days
in green, and T3 = T1 + 196 days, in blue. Data (black dots) are from 
Table~\ref{table:matching_times} and the peak of each curve is marked with a small bar.  
Note that each curve is an explicit solution to a
polynomial fit with 4 constraints: 2 flux density points and 2 slopes. 
T3 data have been extrapolated from the measured data. The black dashed curve
corresponds to the non-linear extrapolation shown by the black dashed curve in 
Fig.~\ref{fig:variability}.  Measurement error bars are typically smaller than the points.
} 
\label{fig:spectra}
\end{figure*}

The results are given in
Table~\ref{table:spectral_fits} and the spectra are plotted in Fig.~\ref{fig:spectra}.
It is now clear that {\it the turnover of the spectral index is shifting to lower frequencies with time,
and the peak of the spectrum also declines with time.} The peak frequency shifts from 4.9 GHz to 4.0 GHz to 
3.2 GHz. 

We have carried out some tests to verify the validity of this result.
For example, 
the same trend can be seen in the original data prior to subtracting the disk emission.
If we repeat this exercise prior to subtracting the disk contribution to both flux density and spectral index,
 then $\nu_m$ is altered by less than 1\% for all time stamps, $S_{max}$ for time steps, T1 and T2, increase
by less than 2\% and the peak flux density for time step, T3, increases by 15\%.
Given the error bars in the L-band spectral index, we also repeated the analysis
assuming that $\alpha_L$ has not varied at all (i.e. letting $\overline{\alpha_{L}}\,\approx\,1.1$ for each time stamp), 
then there is a 1\% decrease in the peak frequency,
$\nu_m$, and a 5\% decrease in the peak flux density, $S_{max}$, for the T3 curve only.  
Fitting a curve, where possible, instead of simple straight lines (e.g. using a second order polynomial for the
L-band data of Fig.~\ref{fig:variability}), also
does not alter the above conclusions. 

 The final issue is related to extrapolation/interpolation.
Note that point T1 is measured at L-band and interpolated at C-band; point T2 is measured at C-band and interpolated at
L-band.  Only point T3 is an extrapolation in time.
If we instead make a more modest extrapolation (i.e. T1 + 100 days, or 44 days after the last measured 
C-band point but still an interpolation at L-band), then the trend
is still consistent with what is shown in Fig.~\ref{fig:spectra}, again with a declining peak that is shifting
to lower frequencies with time.  This intermediate curve (not shown) falls 
 between the T2 and T3 curves of  Fig.~\ref{fig:spectra}.

At time T3, the L-band extrapolation is small, both when one considers the elapsed time (32 days after the
last L-band measurement) as well as the modest
implied variations in flux density and spectral index.
It is the C-band flux density that has undergone the largest change as a result of our extrapolation
(Fig.~\ref{fig:variability}). 
 We have made the simplest possible assumption as to the form of its time decay, i.e. linear, arguing that
to impose anything more complex (and arbitrary) is not justified by the data. 

 However, in anticipation of our model results
(see Sect.~\ref{sec:adiabatic})
we can also try a $S_\nu\,\propto\,t^{-5/3}$ dependence (Eqn.~\ref{eqn:surprisingresult}) on the C-band flux density.
The results are shown as the black dashed curves in Figs.~\ref{fig:variability} and ~\ref{fig:spectra}.  
Note that we restrict this curve only to the region from T2 to T3 in which a true extrapolation at C-band
is needed.
 In addition, the curve's time dependence applies to the highly optically thin regime, whereas
there is actually a changing optical depth with time, as explained in Sect.~\ref{sec:adiabatic} (hence the
`disjoint' slopes at T2).  This new extrapolation increases the
C-band T3 flux density by only 3\% from 209 mJy (Table~\ref{table:matching_times}) to 215 mJy
 and its affect on
the behaviour of the curves in Fig.~\ref{fig:spectra} is minor.

 Any other more dramatic extrapolation
is both unknown and speculative and, as stated above, even modest extrapolations still follow the trends shown in
Fig.~\ref{fig:spectra}.
In summary, we conclude that the behaviour of declining flux densities and frequencies with time is a robust conclusion within the
constraints of the available data.

Fitting the spectrum with a polynomial is the best and most accurate way of describing the source's variable behaviour with time, as
observed.
It does not assume any physical model, however, an issue that we will address in Sect.~\ref{sec:discussion}.


\begin{deluxetable}{lcccccc}
\tabletypesize{\scriptsize}
\tablecaption{Parameters of Polynomial Spectral Fits\label{table:spectral_fits}}
\tablewidth{0pt}
\tablehead{
\colhead{Date} & \colhead{$a_0$} & \colhead{$a_1$} &\colhead{$a_2$}
& \colhead{$a_3$} & \colhead{$\nu_m$\tablenotemark{a}} &\colhead{$S_{max}$\tablenotemark{b}}  \\  
 }
\startdata
Time 1 (T1 = 30-Dec-2011) & -100.27 &  249.79 &  -35.17 &  1.305 &  4.87 &  432.7\\
Time 2 (T2 = 24-Feb-2012)& -156.93 &  338.22 &  -65.22 &  3.845 &  4.03 &  398.5\\
Time 3 (T3 = 13-Jul-2012) & -36.78 &  230.71 &  -53.02 &  3.566 &  3.22 &  275.4\\
\enddata
\tablecomments{$\,$ Parameters of the polynomial fit to the spectrum for 3 different time stamps.
The fit is to $S_\nu\,=\,a_0\,+\,a_1\,\nu\,+\,a_2\,\nu^2\,+\,a_3\,\nu^3$, where $S_\nu$ is the flux density
(mJy) at frequency, $\nu$ (GHz), and simultaneously matching the slope of this function to the
slope of the power law spectrum, $S_\nu\,\propto\,\nu^\alpha$ in each band (see also Sect.~\ref{sec:spectra}).
 Input data are from Table~\ref{table:matching_times}.
}
\tablenotetext{a}{Frequency of the spectral peak (GHz).}
\tablenotetext{b}{Flux density at $\nu_m$ (mJy).}
\end{deluxetable}

\subsubsection{The Absence of Large-Scale Disk Emission} \label{sec:no-disk}

{\it No extended disk emission} beyond the 1.8 kpc diameter central disk seen in
Figs~\ref{fig:optical_Lband_contours} and \ref{fig:optical_Cband_contours} (B/L and C/C data, respectively)
 has been detected from NGC~4845.  This does not appear to be a sensitivity or dynamic range problem when
compared to other galaxies in the CHANG-ES sample \citep{wie15}.  
 In addition, measurable emission is seen in NGC4845 throughout the large-scale disk in all WISE (Wide-field Infrared Survey Explorer; IR)
bands as well as in GALEX (Galaxy Evolution Explorer; UV) images.
Previous radio continuum observations
(Table~\ref{table:image_parameters}) similarly reveal radio emission that is spatially confined to the central
region of the galaxy.
 A thorough discussion of this issue, for example, a study of SF rates and timescales
in comparison to the radio emission, must await a future paper.  We do note, though, 
that truncated gaseous
disks and galaxy stripping are well-known properties of some Virgo Cluster galaxies
\citep[e.g.][]{cay90,ken99,ken04,vol13,mur09}.

%

\subsection{Polarized Emission} 
\label{sec:polarization}

Flux density measurements for linear polarization, $P$, circular polarization, 
$V$, and their fractions
of the total unpolarized flux density, $S$, are given in Table~\ref{table:flux_densities}.
In all arrays and frequencies, the percentage linear polarization is less than or of order 0.5\%. 
Since the polarization calibration cannot be guaranteed to be better than 0.5\%
(S. Myers, private communication)\footnote{See also
the NRAO polarimetry on-line documents.}, we cannot claim to have detected any linear polarization
in any configuration/band combination for NGC~4845.  

On the other hand, there is a clear measurement of circular polarization at L-band, though not
at C-band.  The circular polarization is quite strong, with a flux density of 5.5 to 6.6 mJy,
(cf. the rms noise of 15 to 28 $\mu$Jy beam$^{-1}$; Table~\ref{table:image_parameters}),
 or a 
polarization percentage of 2 to 3\%. Significantly, each independent observation has resulted in
approximately the same percentage circular polarization.
 
It is important to consider whether the circularly polarized emission could have been artificially
introduced.  For example, any errors will affect a $V$ signal more strongly than total intensity
 since $V$ is measuring a {\it difference} between correlated 
signals\footnote{$V\,=\,(RR\,-\,LL)/2$, where $RR$ refers to the correlation of `right-right' -hand
polarization and 
$LL$ refers to `left-left'.}.  It is well known that the VLA suffers from `beam squint'
which can result in spurious signals off-axis \citep{bri03}. 
The center of NGC~4845, however, was placed at the field center where beam squint is zero. 
Self-calibration can also result in spurious $V$ signals if there are other off-axis 
circularly polarized sources in the field
 \citep[see][for details of a related example]{hom99}. 

Our best counterexample, 
however, is the B/L data for which negligible signal was observed other than in the centre of the map. 
This data set was re-imaged with no self-calibration table applied and (although beset
with higher residual sidelobes due to the lack of self-calibration) the resulting image
still shows a significant $V$ signal.  An alternative self-calibration method was also employed\footnote{A `T' type
self-calibration in which a single solution for both polarizations is obtained.}; again, the result was poorer with
higher sidelobes, but there was no change in the V flux density.  Imaging of the primary calibrator, 3C286,
showed $V/I<0.17\%$ and the polarization calibrator, $OQ208$, had  $V/I<0.08\%$.


In summary, the circular polarization appears to be of order a few percent in $V/S$
 and also $V/S_{AGN}$ 
(Table~\ref{table:flux_densities}). Moreover, we find the {\it V spectral index, $\overline{\alpha_V}$}, again 
possible to determine because of our broad bands
(Sect.~\ref{sec:obs_data}),
 to be quite steep, of order $-2\,\to\,-3.5$
(Table~\ref{table:spectral_indices}).  Given the above discussion, we do not claim that the small
variations between arrays are significant.  
The rather striking result of measurable circular polarization and its variation with frequency
will be discussed further in Sect.~\ref{sec:circular_polarization}.

\section{Discussion} \label{sec:discussion}

In this section, we seek to explain the following observations:  a) the turnover in the radio spectrum and its evolution with time
(both amplitude and frequency shifts),
conveniently summarized for this `Gigahertz-Peaked Spectrum' (GPS) source in  Fig.~\ref{fig:spectra},
and b) the detection of circular polarization while a believable linearly polarized signal is absent.  We also
attempt to connect the X-ray observations of \citet{NW13} with our radio data. 

\begin{figure*}[!ht]
   \centering
   \includegraphics*[width=0.5\textwidth]{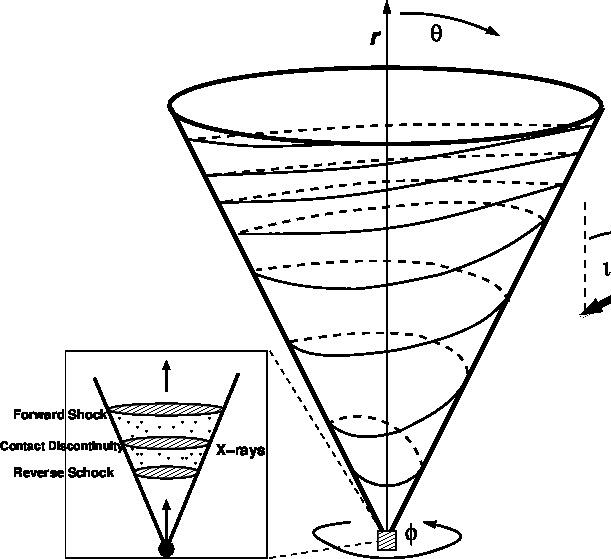}
   \hspace{-1.20in}
   \caption{One side of an outflow, of order 0.1 pc in size, which forms a wide-angle cone 
($\theta$ large) or narrow angle jet (small $\theta$)
of relativistic electrons.
The origin of the outflow is at the apex and the helical curve represents the azimuthal
magnetic field. A suggested line of sight (los), which passes through a sheath of material in or around the cone,
 is shown by the heavy back arrow which makes an angle, $\iota$, with
the radial (along the cone axis) direction. A depolarizing screen (not shown) is outside of the cone along the los.
{\it Inset:} Blow-up of the apex region, of order 1000 AU in size, showing the shocked region in which X-rays are
generated.}
\label{fig:cone}
\end{figure*}

We describe more generally 
a jet/cone model with basic geometry that lies behind our general thinking in Appendix~\ref{appendixA}. 
For this section, it is sufficient to imagine a source of relativistic particles as illustrated
in the cone geometry of  Fig.~\ref{fig:cone}.  The 
optical cone observed in Fig.~\ref{fig:optical_Cband_contours} suggests such a geometry on
larger scales, possibly representing an earlier outflow. As we do not know the small-scale opening angle, 
however, we will use `jet' or 'cone' interchangeably when referring to our model.

We are aided by the fact that
the X-ray outburst was observed and provides us with a timescale.  We will refer to the
time of the peak of the X-ray light curve (22-Jan-2011) as $t_0$ 
\citep[the disruption of a mass that apparently triggered
the X-ray flare occurred 60 to 100 days earlier,][]{NW13}. Our radio data span 6 months
(Table~\ref{table:image_parameters}) and our first timestamp from our spectral fits 
(Table~\ref{table:spectral_fits}) occurs at T1 = 342 days $\approx$ 1 year after
the X-ray peak.

Several reasonable simplifications can be made.  First, for a very fast jet, transformations would be
required between the source and observer's rest frame for parameters such as jet angle, brightness, magnetic field, etc.
For completeness, these transformations are provided in
Appendix~\ref{appendixB}; however, we find that
unless the bulk jet speed is ultra-relativistic, such transformations
are not required and will not be applied here.  
Secondly, our calculations are order
of magnitude estimates to test for feasibility and consistency with the observations.


\subsection{Source Size}
\label{sec:size}

From observations, the core of NGC~4845 is unresolved, likely less than 0.5 arcsec (Sect.~\ref{sec:astrometry}) so
we do not have sufficient resolution to measure its size, $\theta_s$, directly.  However, the variability itself suggests
a limiting size.  If the source was created with the X-ray flare observed by INTEGRAL in January, 2011
\citep{NW13}, then the projected physical size should be 
~$\le ct_0\sin{(\iota)}\approx 10^{18}\sin{(\iota)}$ cm, where $\iota$ is the inclination of the major source axis to the line of sight
(Fig.~\ref{fig:cone})
and $t_0$ is the elapsed time.  For $t_0\approx 1$ year (cf. dates of Table~\ref{table:image_parameters}) and a distance of
 $17$ Mpc (Sect.~\ref{sec:introduction}),
  this suggests that $\theta_s \le 4 \sin{(\iota)}$ mas.
If $\iota=0^\circ$, then, we are looking down the major axis of the cone along the outflow
 and $\theta_s$ would relate to the cross-sectional diameter of the cone;
if  $\iota=90^\circ$, then the cone is aligned with the sky plane and $\theta_s$ would relate to a typical 
cone size in projection.
In the following, we adopt $\theta_s \sim 1$ mas ($\sim$ 0.1 pc or $\approx\,10^{17}$ cm) as a fiducial value.

Note that the brightness temperature at $20$ cm 
is $\approx 3.4\times 10^{11}~K$ if the source size is of order 1 mas.
 This is still below the limit for catastrophic inverse Compton losses. 
GPS sources at higher redshift suggest a range of sizes with the lowest $\approx$ 1 pc \citep{fan09}, hence 
the core of NGC~4845 extends the array of GPS source sizes an order of magnitude smaller.

\subsection{The Low Frequency Spectral Turnover}
\label{sec:lowfreqturnover}

A consistently difficult problem in the past has been to understand whether low frequency turnovers are caused
by thermal absorption or synchrotron self-absorption (SSA).  In this section we explore both possibilities and
conclude that the dominant cause of the turn-over is likely synchrotron self-absorption.

We first, however, exclude Razin 
suppression of the low frequency emission 
as the cause of the turnover
because of the predicted exponential cutoff below the 
Razin frequency \citep{Mel80} that is not observed. 
At least for a homogeneous source, the effect also does not give a dependence of circular polarization on 
frequency which in fact we detect (Sect.~\ref{sec:polarization}).

\subsubsection{Synchrotron Self-Absorption}
\label{sec:ssa}

For synchrotron self-absorption (SSA) as an explanation for a GPS source \citep[see e.g.][]{ACT08} 
one expects the brightness temperature of the source to approach the electron kinetic temperature \citep[e.g.][]{Long94}. 
The source brightness temperature in terms of the wavelength, $\lambda$, and flux density, $S_\nu$, is 
\be
T_b=\frac{\lambda^2}{2k}\frac{S_\nu}{\Omega_s}~K\approx \frac{7\times 10^{7}\lambda_6^2}{\theta_s(mas)^2}~S_\nu~K,\label{eq:Tb}
\ee
where $\Omega_s=(\pi/4)\theta_s^2$, $\theta_s$ is the source size, and $k$ is Boltzmann's constant. 
We have taken $\lambda$ in units of $6$ cm, $\theta_s$ in mas and the flux density in $mJy$.
The kinetic temperature, $T_e$, for 
relativistic electrons is given by $kT_e=\gamma m_e c^2/3$.
Equating these two temperatures gives the limiting self-absorbed flux density at a peak frequency for electrons of energy,
$\gamma$, as 
\be \label{eqn:S_selfabs}
S_{\nu}=\sqrt{\frac{2}{3}}\frac{\pi}{3}\frac{m_e\theta_s^2}{\nu_g^{1/2}}~\nu^{5/2}\approx 950~ \frac{\nu_5^{5/2}\theta_s(mas)^2}{B_\perp(-2)^{1/2}}~~mJy.
 \ee 
where $\nu_g\equiv eB_\perp/(2\pi)m_ec$ is the gyrofrequency, $e$ and $m_e$ being electronic properties, and $\nu_5$ is the frequency 
in units of 5 GHz.
From this last expression we can infer the universal (thermodynamic) self-absorbed spectral index, $\alpha=2.5$
and declining flux densities with decreasing frequency.

Adopting $B_\perp(-2)\,=\,1$ (see Appendix~\ref{appendixC})
and  $\theta_s(mas)=1$ (Sect.~\ref{sec:size}), the observed flux density at $6$ cm is lower than the flux density given by
Eqn.~\ref{eqn:S_selfabs}, indicating that self-absorption is not complete at this frequency.
However, at $20$ cm, the observed flux density is much higher than that of Eqn.~\ref{eqn:S_selfabs} (by an order of magnitude).  
In other words, the observed flux density at $6$ cm implies a peak (turnover) frequency below $6$ cm whereas the
observed flux density at $20$ cm implies a peak frequency well above $20$ cm.
Thus we expect our results summarized in Fig.~\ref{fig:spectra} to describe a spectral region that is 
transiting towards self-absorption at the low frequencies.          

\begin{figure}[!ht]
   \centering
   \includegraphics*[width=0.4\textwidth]{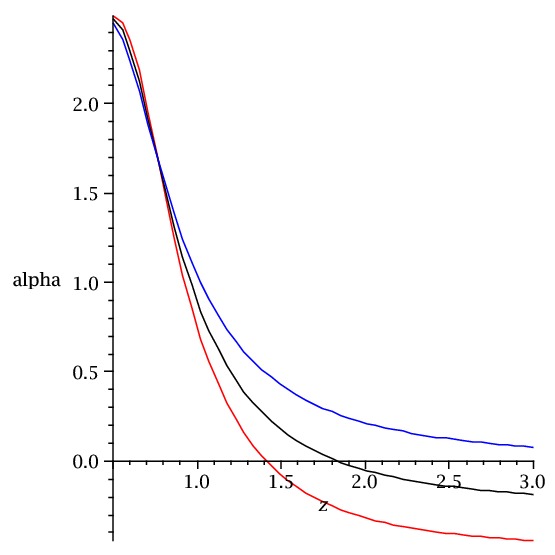}
   \caption{The figure shows the rollover in the spectral index during the onset of synchrotron self-absorption at low frequencies. 
The frequency is measured in units of $\nu_1$, the frequency at which the optical depth is 1
(i.e. $z\,=\,\nu/\nu_1$). 
This is $0.707$ times the peak frequency for $p=2$. 
The lower curve at large frequency is our reference case for which $p=2$ ($\alpha\,=\,-0.5$). 
The higher curves at large frequency taken in rising order are for $p=1.5$ and $p=1$ respectively. The transition from thin to moderately thick is over roughly a factor of three in frequency. Another factor of two in frequency is required to attain the completely self-absorped index.}
\label{fig:alpharollover}
\end{figure}
\begin{figure*}[!ht]
   \centering
   \includegraphics*[width=0.49\textwidth]{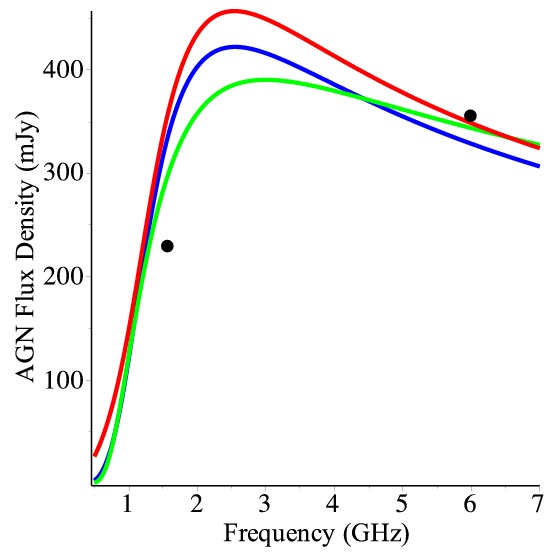}
   \includegraphics*[width=3.2in]{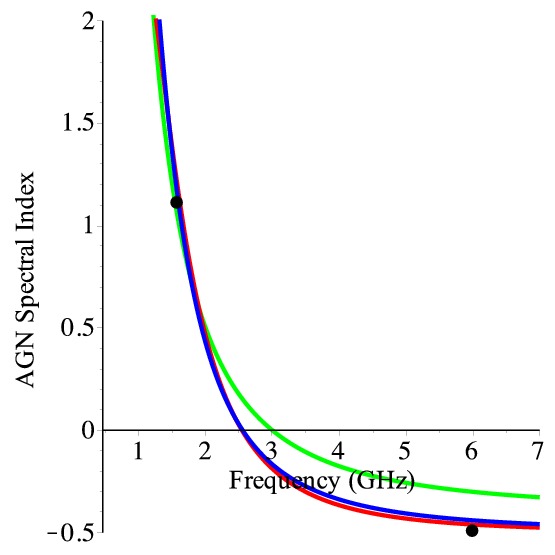}
   \hspace{-0.20in}
   \caption{Best fit synchrotron-self-absorption-only spectrum (red), thermal-absorption-only (green) 
and a synchrotron self-absorption plus a thermally absorbing screen (blue), all as applicable to time, T2.
Best fit parameters are provided in Table~\ref{table:synchthermal_fits}. Black dots represent the data.
The left plot shows the spectra, displaying fits to the flux densities, whereas the right plot shows the slopes, displaying
fits to the spectral indices.  Measurement error bars are typically smaller than the points.
} 
\label{fig:therm_synch}
\end{figure*}

One of the principal results of these observations is the measurement of the in-band spectral indices, 
which have been used to define the fits of Fig.~\ref{fig:spectra}. If we are to explain these curves as a 
transition to synchrotron self-absorption, then we must consider the roll-over in spectral index from an
optically thin spectral index of $\alpha=(1-p)/2\,\approx\,-0.5$ $=> p = 2$ (Table~\ref{table:spectral_indices})
~(using~ $N(E)=N_oE^{-p}$~ for the energy distribution 
function of the electrons) towards the index $\alpha=2.5$ which is expected when the source is fully self-absorbed, as shown above. 
This roll-over is described in \cite[][p. 97]{Pach70}. We calculate 
\be
\alpha=\frac{d\ln{J(z)}}{d\ln{z}},
\ee
where
\be
\label{eqn:Jzequation}
J(z)=z^{5/2}\left(1-\exp{(-z^{-(2+p/2)})}\right),
\ee
and ~$z\equiv\nu/\nu_1$. The frequency at which the optical depth equals one is ~$\nu_1$, which for $p=2$ is given 
by ~$\nu_1=0.707 \nu_c$, $\nu_c$ the critical frequency. The result is shown in Fig.~\ref{fig:alpharollover}
for several values of $p$. The reference case for our purposes is the lower curve at large $z$ for which $p=2$.
We see that $\alpha$ varies from the optically thin value of $-0.5$ to a moderately optically thick value of 
$\alpha =+1.0$ (as observed at L-band) over roughly a factor three in frequency. However because of the steepness of the rising curve at ~$z=1$~ 
it is possible to have $\alpha$ in the optically thick region ranging from just below $0.5$ to $1.0$ over 
essentially the same range in frequency. This is a 
reasonable explanation of the curves in Fig.~\ref{fig:spectra} between $6$ and $20$ cm (C-band and L-band,
respectively).  Given the index $p$, we emphasize that the only free parameter is $\nu_1$ in
describing the spectral shape.  To detect the 
fully self-absorbed spectral index, one has to descend another factor of two in frequency.  A fit by the synchrotron model to all the data is included in Fig.~\ref{fig:therm_synch}.

We can also estimate the relativistic electron density required to produce the turnover.
The absorption coefficient for the synchrotron particles is approximately 
\be
{\kappa_{er}}_\nu=1.7\times 10^{-14}N_o\frac{B_\perp(-2)^2}{\nu_1^3}~cm^{-1},
\ee
where the subscript, $er$, refers to relativistic electrons and $\nu_1$ is in units of GHz. 
When $p=2$,~the coefficient ~$N_o\approx n_{er}E_o$. Here $n_{er}$ is the relativistic electron density and $E_o=\gamma_om_e c^2$ is the lower energy cutoff of the distribution. If $L$ measures the line of sight distance through the source then we conclude,
 by setting the optical depth equal to one that
\be 
n_{er}~L=2\times 10^{19}\frac{1}{\gamma_o(10)B_\perp(-2)^2}~cm^{-2}.
\ee
Here ~$\gamma_o(10)$ is the lower cutoff in units of ten (which we set to 1, following other authors)
and we have taken ~$\nu_1=1.5$. Nominally this suggests, 
given our proposed source size ($10^{17}$)
that $n_{er}\,\approx\, 200$ $cm^{-3}$.  

If the number of protons is the same, then a rough calculation of mass for a cylinder (approximating
a cone) of length and diameter, $10^{17}$ cm, would result in $\sim\,3\,\times\,10^{29}$ g (more if
thermal gas is entrained). This is 5 to 10\% of the mass of
$3\times 10^{30}\le M\le 6\times 10^{30}$ g estimated by \cite{NW13} to have produced the X-ray flare via 
accretion of a `super-Jupiter' object.
 In the latter case, the authors suggest that 10\% of a `super-Jupiter' of
mass, 14 - 30 Jupiter masses, has produced the X-ray flare.  This suggests some consistency between 
the relativistic electron density required to explain the radio spectral turnover and the expected quantity of
particles supposed to be involved in the X-ray event (Fig.~\ref{fig:cone}).

 For a more thorough investigation of SSA as a possible cause of the turn-over, we have also fit the T2 timestamp
with a pure SSA spectrum, allowing $\nu_1$, $p$, and the amplitude
to freely vary.  For the latter, we introduce a scaling factor, $\tilde C$, which converts the amplitudes to
a flux density, $S(z)$, for comparison to the data, i.e.  $S(z)\,=\,\tilde C\,J(z)$.  The best fit is chosen by 
minimizing $\chi^2$ during a least squares fit.  

The resulting best fit parameters are given in Table~\ref{table:synchthermal_fits} and the spectrum, along with its derivative, are
plotted as the red curves in Fig.~\ref{fig:therm_synch}. 
This curve is identical to the generic red curve in Fig.~\ref{fig:alpharollover} when $\nu_1=1.8~ GHz$.

\begin{deluxetable}{lccccccccc}
\tabletypesize{\scriptsize}
\tablecaption{Parameters of Synchrotron and Thermal Spectral Fits at T2\label{table:synchthermal_fits}}
\tablewidth{0pt}
\tablehead{
\colhead{ } & \colhead{$p$} & \colhead{$\nu_1$} &\colhead{$\tilde\kappa$}
& \colhead{$\tilde C$} & \colhead{$S_L$} &\colhead{$S_C$} &\colhead{$\alpha_L$} &\colhead{$\alpha_C$} &\colhead{$\chi^2$} \\  
 }
\startdata
Synchrotron\tablenotemark{a} & 2.0 & 1.8  &  & 645  & 358  & 349 & 1.20 & -0.460 & 118 \\
Thermal\tablenotemark{b} & 1.8 &  & 1.8 & 740  & 299  & 344 & 1.05 & -0.300 & 249 \\
Synchrotron + Thermal\tablenotemark{c} & 2.0 & 1.5  & 0.3 & 675 & 336  & 329  &1.16  &-0.439 & 117 \\
Data\tablenotemark{d}  &                     &      &     &     & 229 &    355    & 1.11 & -0.493 \\
\enddata
\tablecomments{$\,$ Best fit parameters for the various fits as described in Sects.~\ref{sec:lowfreqturnover} and
Fig.~\ref{fig:therm_synch}. Subscripts, $L$ and $C$ refer to L-band and C-band, respectively. $\chi^2$ is a measure of
the goodness of fit, here defined as 
$\chi^2= \{[S_L(fit)-S_L(data)]^2+ [S_C(fit)-S_C(data)]^2\}/(\Delta S)^2 +
\{[\alpha_L(fit)-\alpha_L(data)]^2+ [\alpha_C(fit)-\alpha_C(data)]^2\}/(\Delta \alpha)^2$, where we have taken
$\Delta S = 5$ mJy and $\Delta \alpha = 0.01$ as representative values. Note that $\tilde C$ is defined differently,
depending on the fit (see Sect.~\ref{sec:lowfreqturnover}).
}
\tablenotetext{a}{$\,$Low-frequency turnover due to synchrotron self-absorption only. }
\tablenotetext{b}{Turnover due to thermal absorption only.}
\tablenotetext{c}{Turnover due to synchrotron self-absorption plus a foreground thermal screen.}
\tablenotetext{d}{Data are from Table~\ref{table:matching_times}.}.
\end{deluxetable}

\subsubsection{Thermal absorption}

 In this section we consider the synchrotron emission to be optically thin and  the turnover to be due to
thermal absorption only.

 In the simplest approach, since
we know the optically thin spectral index, which we take to be $\alpha\,\approx\,-0.5$ (Table~\ref{table:spectral_indices}), 
the spectral
index in the optically thick ($\tau\,=\,1$) regime that would result from thermal absorption can be computed.
In Appendix~\ref{sec:thermal_absorption} we show that purely thermal absorption can not provide a satisfactory fit to both the low frequency and high frequency spectral indices. If one improves the fit at one end by adjusting the optical depth, it becomes much poorer at the other end. 


Just as for the synchrotron model, a more systematic treatment of thermal absorption is
to solve for the spectrum as well as its derivative by finding the parameters of best fit.
To achieve this, we also allow the synchrotron energy spectral index, $p$, to vary freely which, as before, fixes the
 optically thin synchrotron spectral index, $\alpha\,=\,(1-p)/2$.  The amplitude scaling,
 $\tilde C$, incorporates $C$ of Eqn.~\ref{eqn:thermalabsorption} as well as a term to convert from specific
intensity to
flux density for comparison to observations.  Finally, we combine the remaining unknown parameters into
a single free parameter, $\tilde\kappa$, which is a variant of Eqn.~\ref{eqn:kappa_thermal} that also includes the 
line of sight distance,
$s_{17}$ (in units of $10^{17}$ cm), and the frequency at which the synchrotron spectrum becomes optically thick, 
$\nu_1(9)$, in GHz. That is 
\begin{equation}
\tilde\kappa\,=\,\frac{{n_e(4)}^2\,s_{17}}{{\nu_1(9)}^2\,T_4^{3/2}}
\end{equation}
where $n_{ec}(4)$ is the thermal electron density in units of $10^4$ cm$^{-3}$ and $T_4$ is the `cold' electron temperature
in units of $10^4$ K.

The resulting best fit parameters are listed in Table~\ref{table:synchthermal_fits} and are shown as green curves
in Fig.~\ref{fig:therm_synch}. This fit is demonstrably poorer than that of pure SSA,
consistent with our statements above. One notes particularly the poor fit to the spectral index in C band.

\subsubsection{Synchrotron spectrum with a foreground thermal screen}





It is natural to imagine thermal absorption ocurring in a sheath of mixed cold and hot particles surrounding the synchrotron jet, given the lack of linear polarization that we observe (see Appendix \ref{appendixC}). In principle this could be a dynamic sheath of cold gas entrained (that is mixed) with the jet particles, or it could be be a stationary surrounding cloud. 

We continue to assume that the synchrotron emitting particles are additionally self-absorbed. Then the equations of the fit are 
\bea
S_\nu&=&\tilde C z^{2.5}(1-\exp{(-\frac{1}{z^q})})\exp{(-\frac{\tilde\kappa}{z^2})},\nonumber\\
\alpha&=& 2.5+\frac{2\tilde\kappa}{z^2}-\frac{q}{z^q}\frac{\exp{(-\frac{1}{z^q})}}{(1-\exp{(-\frac{1}{z^q})}}.\label{eq:synchplustherm}
\eea
We take $q\equiv (2+p/2)$, $z= \nu/\nu_1$ as before, and $\tilde\kappa$ as above.
This model has four free parameters ~$p,\nu_1, \tilde\kappa$~ and the amplitude ~$\tilde C$. 
The results are given in table \ref{table:synchthermal_fits} and by the blue curve in Fig.~\ref{fig:therm_synch}.

The fit is no better than that of SSA alone although it is equally good. The high frequency spectral index problem of the pure thermal absorption has been removed. However this has been accomplished by removing almost completely the thermal absorption at C band ($\propto e^{-\frac{0.3}{16}}$), which leaves the optically thin synchrotron emission unchanged there. Nevertheless this is an acceptable configuration. 
  
We observe also from Fig.~\ref{fig:therm_synch} that any of these simple models fails to reproduce the fluxes to observational accuracy. This is not too surprising since we have ignored any real geometrical structure in the source. For example we see from the definition of ~$\tilde\kappa$~ that the line of sight required for a given absorption is rather sensitive to $T_4$. Moreover a detailed radiative transfer calculation in the 
style of \cite{J88} for a given (partially optically thick) geometry may be required.

It is interesting to note that \cite{WJ-Z14} have invoked very similar ideas to our own regarding SSA and SSC (Synchrotron Self-Compton) in the description of a more distant TDE (Sw J16449.3+573451). Their geometry requires two nested conical jets of different speeds and total energies. The slower outer jet may be compared to our idea of a dynamic sheath. However we differ from these authors in using the jet to produce the early X-ray 
emission in a shock lobe followed by the radio emission after the decelerated jet has expanded. 

A detailed study is justified for their source since much more data in the x-ray and radio is available than is the case for NGC~4845. For this reason we will adopt our simplest model with the smallest number of parameters, namely pure SSA, in the next section on variability. Subsequent more detailed work after future VLBI obervations may be justified.





\subsection{Variability of the Spectrum}
\label{sec:adiabatic}

\subsubsection{A Single Injection of particles followed by fading does not fit the data}
In a paper discussing a distant GPS source \citep{ori10} it was concluded that the 
overall source spectrum was best explained by an ensemble of electrons in which acceleration had ceased. 
The authors called this the `fader' phase. Their extensively sampled spectra were fit in terms of an 
`injection spectral index' and in terms of the ratio of the acceleration cutoff time to the synchrotron lifetime at peak frequency. 

In the most successful fit to the overall spectrum (that is with a reasonable injection spectral index of
 $\alpha\,\approx -0.7$) they found a
cutoff time which was about~ $20\%$~ of the synchrotron lifetime of several thousand years. However these authors 
did not observe a fresh outburst in this source so the expected dependence on time could not be checked. 
This is the distinct advantage of detecting the combined X-ray and radio emission of the outburst in NGC 4845. 

We show in Appendix~\ref{sec:simple} that the time evolution of the spectra shown in Fig.~\ref{fig:spectra} does not
match this simple picture.

\subsubsection{Spectral Variability as a Result of Adiabatic Expansion}
\label{sec:adiatic}

We now wish 
to understand the time evolution of the source spectrum, again summarized in Fig.~\ref{fig:spectra}
(from epochs $T_1$ to $T_3$),  which we take to be the best representation of the data. As remarked above, we assume SSA uniquely since this minimizes the number of free parameters and provides a fit to the observations which is equally as good as a synchrotron spectrum with an
additional thermal screen.
We therefore suppose the energy spectrum, $N(E)$, of the relativistic particles to be evolving due to  energy loss through adiabatic expansion.
 
This mechanism presupposes pressure equilibrium with the medium surrounding the evolving source, against which the 
relativistic particles are working. This medium may be thermal gas in a `sheath' around the source, perhaps
 combined with (or indeed dominated by) the `hoop stress' of a helical magnetic field. Such a local confining 
magnetic field must be created by an externally driven current.  

During adiabatic expansion the equation for the total electron density per unit volume and per unit energy,
 $n({\bf r},E,t)$, is (ignoring acceleration, diffusion and particle source terms)
\be
\partial_tn+\nabla\cdot({n\bf v})=\frac{1}{3}(\nabla\cdot{\bf v})\partial_E(nE).
\ee
where {\bf r} is a position vector, {\bf v} is a velocity, and $t$ is time.
For our purposes we integrate this over an expanding volume that contains the same total number of particles at any given time, 
in order to get the familiar `diffusion-loss' equation \citep[e.g.][]{Long94} for the number  of particles per
unit energy interval, $N$
\be
\partial_tN(E,t)=\frac{1}{3}(\nabla\cdot{\bf v})\partial_E(NE).
\ee
We have assumed that the velocity divergence can be expressed wholly as a function of time (including a constant) in order to perform the spatial integration. 

For simplicity (but see Appendix~\ref{appendixA}) we assume that the source expands with the uniform radial velocity, 
$v_o\widehat{\bf e}_r$, where the radial direction, $r$, is along the axis of the cone. 
Hence ~$\nabla\cdot {\bf v}=2v_o/r=2/t$~, since $r=v_ot$. The solution for $N(E,t)$  is then found (e.g. by the method of characteristics) to be
\be
N(E,t)=N_o E^{-p}\left(\frac{t}{t_1}\right)^{(2(1-p)/3)}.\label{eq:N}
\ee 
Here, the reference time, $t_1\,=\,342$ days (elapsed time since the X-ray peak at $t_0$) occurs at the time stamp, $T1$.
The energy spectrum at time, $t_1$, has clearly been taken as usual to be $N(E)=N_o E^{-p}$. Subsequently the spectrum is unchanged
 but now, when $p=2$, 
\be\label{eqn:N_now}
 N_o(t)=N_o(t_1)(t/t_1)^{-2/3} 
\ee

The frequency at synchrotron optical depth one ~$\nu_1$~ (which is proportional to the peak frequency, $\nu_m$) satisfies \citep{Pach70} 
\be
\label{eqn:nu1synch}
\nu_1\propto N_o(t)^{1/3}B_\perp ^{2/3} s^{1/3},
\ee
when $p=2$. If we suppose that the line of sight size, ~$s\propto t$ (as does $r$), and use the form ~$N_o(t)$~ found above, then 
\be\label{eqn:peakfreq}
\nu_1\propto (t/t_1)^{1/9}B_\perp^{2/3}.
\ee

We would like to explain the ratios of the peak frequencies 
 which, from 
Table~\ref{table:spectral_fits}, are 
$ \nu_m(T1)/\nu_m(T2)=4.87/4.03\approx 1.2$ and $ \nu_m(T1)/\nu_m(T3)=4.87/3.22\approx 1.5$
in terms of this expansion. 
If $B_\perp\propto 1/s$ so that the azimuthal field dominates (the azimuthal field wraps around the cone/jet, Fig.~\ref{fig:cone}),
then we calculate the ratios
 as $(1+56/342)^{5/9}\approx 1.1$ and $(1+196/342)^{5/9}\approx 1.3$, respectively.
 If, at the other extreme, we suppose that the radial field dominates $B_\perp$,
then ~$B_\perp\propto 1/s^2$ and we calculate the 
ratios as $(1+56/342)^{11/9}\approx 1.2$ and $(1+196/342)^{11/9}\approx 1.7$, 
respectively.  
Overall, either decline in magnetic field approximately fits the data; however, we show below that  
$B_\perp\propto 1/s^2$ is required to explain the amplitude variations.
This suggests a rather losely wound, conical magnetic helix together with moderate particle pitch angles in the source.

Finally, we wish to consider the changing amplitudes of these peaks.  The specific intensity at the peak frequency, $\nu_m$, is given by
\citep[][p. 97]{Pach70}
\be
I_{\nu_m}\,\propto\,{B_\perp}^{-1/2}\,\nu_1^{5/2}
\ee
Applying Eqn.~\ref{eqn:peakfreq}, this becomes
\be
I_{\nu_m}\,\propto\,\left(N_0(t)\right)^{5/6}{B_\perp}^{7/6}\,s^{5/6}\,\propto\,t^{-37/18}
\ee
The latter time dependence results from the fact that each of the above terms can be expressed as a function of time using
Eqn.~\ref{eqn:N_now},  $B_\perp\propto 1/s^2$ from above, and taking
a constant velocity so that $s\propto t$.  

It remains to convert from specific intensity to the observed flux density which requires considering how the projected area of the source
varies with time.  For spherical outflow, this dependence would be ${\theta_s}^2\,\propto\,s^2\,\propto\,t^2$.  However, a jet/cone
geometry should have $v_r\,>> \,v_\perp$, where $v_r$, $v_\perp$ are velocities along the axis of the cone (the radial direction) and
perpendicular to it, respectively, in which case  ${\theta_s}^2\,\propto\,s\,\propto\,t$.  In the 
unlikely event that $\iota\,=\,0$ (the
line of sight is directly along the axis, Fig.~\ref{fig:cone}), then ${\theta_s}^2\,\approx\,$ constant
for a narrow outflow.  Adopting the jet/cone geometry, then the time dependence for
the peak flux density is finally
\be
\label{eqn:snumax}
S_{\nu_{m}}\,=\,{\theta_s}^2 I_{\nu_m}\,\propto\,t^{-19/18} \approx\,1/t
\ee 

Again, from Table~\ref{table:spectral_fits}, the amplitudes of the peaks at time stamps, $T1$ to $T3$, 
are in the ratio, 
 $S_{T2}/S_{T1}\,=\,398.5/432.7\,=\,0.92$, and
 $S_{T3}/S_{T1}\,=\,275.4/432.7\,=\,0.64$,
 and the predicted ratios are,
$(1+56/342)^{-1} = 0.86$ and
$(1+196/342)^{-1} = 0.64$, i.e. in agreement to within 7\% or better.


If we repeat the above, assuming that  $B_\perp\propto 1/s$, then the result is $S_{\nu_{m}}\propto t^{1/9} $ which would 
result in an {\it increasing} flux density with time which is clearly not observed.  Thus, hereafter, we adopt
$B_\perp\propto 1/s^2$.

For the sake of completeness in the context of our model, we can also find the time 
dependence of the flux density, $S_{\nu}$, for
any fixed frequency, $\nu$, in the optically thin limit (as opposed to
$S_{\nu_m}$, since ${\nu_m}$ changes with time as found above).
For this case,
\be
S_{\nu}\,=\,{\theta_s}^2\,I_{\nu}\,=\,{\theta_s}^2\,I_{\nu_1}\,J(z)
\ee
where $J(z)$ is given by Eqn.~\ref{eqn:Jzequation} and $I_{\nu_1}$ is related to 
$I_{\nu_m}$ by a simple constant.  The time dependence of ${\theta_s}^2\,I_{\nu_m}$
is given by Eqn.~\ref{eqn:snumax} and, in the optically thin limit
with $p=2$, $J(z)\,\propto\,{\nu_1}^{1/2}$.
Since $\nu_1\,\propto\,t^{-11/9}$ as already found from Eqn~\ref{eqn:peakfreq} 
with $B_\perp\,\propto 1/s^2$ and $s\,\propto\,t$, we find finally
\be
\label{eqn:surprisingresult}
S_{\nu}\,\propto\,t^{-{5/3}}
\ee

Such a dependence might be expected at C-band where the emission is optically thin. Note, however, that
we base our analysis on the (exact) polynomial spectral fits to all of the data (fluxes and slopes) in 
Fig.~\ref{fig:spectra}, rather than on partial spectral fits to the data as displayed in
Fig.~\ref{fig:therm_synch} for T2.  It is readily seen that the shape of the polynomial curves
is not always consistent with the C-band flux density lying in the optically thin region, where
the $t^{-{5/3}}$ decline should apply.  This is because the value of 
$z\,=\,\nu_6/\nu_1$ in C-band for T1 is found to be only 1.74.  Fig.~\ref{fig:alpharollover} then
shows that $\alpha_6\,=\,-0.224$.  For T2, $z\,=\,2.1$ and thus  $\alpha_6\,=\,-0.341$.  That is,
with increasing time, the spectrum becomes more optically thin, consistent with 
what might be expected from adiabatic expansion, but the fully optically thin regime is later
than T2.

Although the T2 polynomial spectrum is still not convincingly optically thin, it is the best
point from which to try the $t^{-{5/3}}$ extrapolation that was carried out for the C-band 
curve in Fig.~\ref{fig:variability} (dashed curve) as an alternative to the
simpler linear extrapolation
(dotted curve).

In summary, both the changing frequencies and amplitudes of the radio spectra are in reasonable agreement with a self-absorbed cone/jet model
which is adiabatically expanding and originates at the time of the hard X-ray flare peak.

\subsection{Circular and Linear Polarization}
\label{sec:circular_polarization}

The most evidently remarkable results concerning polarization are reported in Tables~\ref{table:flux_densities} and 
\ref{table:spectral_indices}. NGC 4845 is in the extension of the Virgo cluster at high Galactic latitude 
($b=+74^\circ$) so these measurements are unlikely to be contaminated by the Milky Way. 

The first table reports significant values (between $2$ and $3$\%) 
of circular polarization (circularly polarized fraction, or cpf) in L-band at all resolutions. These are large compared to maximum values 
previously observed in GPS sources \citep[$\approx 1$\% e.g.][]{OS13}, although not highly 
discordant, given the uncertainties (Sect.~\ref{sec:polarization}). Equally remarkable  
is the non-detection of circular polarization at C-band. There is evidently an extremely rapid variation of 
circular polarization with frequency, reported in the second table, i.e. 
the in-band spectral index of the circularly polarized flux lies between $-2.2$ and $-3.4$.
This behaviour is coupled with no reliable detection of linear polarization.
 We need to constrain the source structure to fit these facts.

A satisfactory explanation of our observed polarization behaviour is difficult to find in the literature.
In this section, we investigate several approaches, and then discuss our proposed explanation which involves
{\it conversion} from linear polarization to circular polarization via {\it generalized Faraday rotation} \citep[see also][]{OS13}.

The intrinsic circular polarization percentage of optically thin synchrotron radiation 
is ~$\cot{\phi}/\gamma_e$, where $\phi$ is the pitch angle and $\gamma_e$ is the electron Lorentz factor corresponding to the peak frequency.
For a  $0.01$ G field, $\gamma_e\approx 200$  in L-band (Eqn.~\ref{eqn:critfreq}), or
 ~$\gamma_e\approx 100$ for a $0.04$ G field. Then 
one would need ~$\phi\,=\,14^\circ \rightarrow 26^\circ$ (for $\gamma_e\approx 200 \rightarrow 100$, respectively)
 in the observer's frame to account for a 2\% circular 
polarization percentage.
Such a pitch angle reduces the emissivity to about $12\rightarrow 30$ \% of the maximum ($\propto (B\sin{\phi})^{3/2}$) which, 
to maintain the observed flux at the same frequency, would require a rather large magnetic field ($10^{-1}$ G) and hence
 magnetic energy ($\sim 10^{49}$ ergs).

The major difficulty with this explanation, however, is that the intrinsic polarization predicts a  
spectral index ~($\overline{\alpha_V}$) for the circularly polarized flux density of ~$-0.5$. This conflicts with what we 
(Table~\ref{table:spectral_indices}) and others \citep[see discussion in][]{OS13} have observed. 
A second difficulty is that we detect little or no linear polarization. This would not be expected for optically thin, homogeneous emission. 
In any case we have already made a case that our frequency observations embrace the transition to self-absorption
(Sect.~\ref{sec:ssa}).
The optically thick regime is also necessary for our proposed explanation by generalized Faraday rotation 
\citep[e.g.][]{Pach73,J88,KM98,ZK2002,M2002,BF2002,OS13}.

\cite{J88}
 gives detailed polarization calculations for a  generalization 
of a definite jet model \citep[see][]{BK79}
and finds circular polarization levels in the simulations comparable to our own. 
Unfortunately the calculated linear polarization  tends to be of the same order, and Jones does not discuss the 
frequency dependence in the optical depth transition zone explicitly. 
His analytic discussion  focuses on the surface where the optical depth is one, whereas we suspect the essential behaviour
to be in transition across this value. The results are complicated by the assumption of 
inhomogeneity comprised of turbulent cells, which is a structure also invoked by \cite{BF2002}.   

In \cite{Pach73},  conversion of linearly to circularly polarized flux  in a source comprised of 
mainly cold electrons with an admixture of relativistic particles is discussed. However $\overline{\alpha}_V=-1$ in this limit. 
Linear polarization is suppressed  only by the average effect of the turbulent cells, if 
the field is indeed randomly distributed and the Faraday rotation per cell is large. When the Faraday rotation per cell is small,
the circular polarization has the~ $\nu^{-3}$ dependence near optical depth unity, but the linear polarization is larger
\citep{BF2002}. This may have to be depolarized in an inhomogeneous, Faraday rotating screen.

\subsubsection{Generalized Faraday Rotation}

Generalized Faraday rotation occurs \citep[e.g.][]{KM98} when the orthogonal normal modes in the 
source plasma are elliptically polarized rather than circularly polarized (effectively linearly polarized) as is the case for standard Faraday rotation. The latter process occurs in a cold (thermal) electron plasma and is proportional to ~$B_\parallel$. The former process requires an admixture (becoming dominant in a relativistic limit) of relativistic electrons and is proportional to ~$B_\perp^2$. Normally the generalized 
process produces a cyclic generation of circularly polarized flux
from linearly polarized flux (this is the `conversion'). That is the case, provided that the plane of linear polarization rotates 
along the line of sight so that there is appropriate rotation on the Poincar\'e sphere \citep{KM98}. Thus, 
either the magnetic field rotates or some cold plasma must be present to Faraday rotate the plane of polarization. The cyclic variation modulates the frequency dependence of the cumulative phase shift between the two modes. If the wave number difference is ~$\Delta k$~ then this phase shift is ~$\Delta k s$, where $s$ is the line of sight distance through the source.

It is important to note that the steep in-band $V$ spectral indices of Table~\ref{table:spectral_indices}
 apply over a relatively narrow frequency range ($500~MHz$) where 
the transition to self-absorption is occurring, as is indicated by the spectral index~$\alpha\approx 1$. In a region where relativistic 
particles dominate \citep{KM98} and optical depth effects are neglected, the percentage cpf, $V/S$, is proportional to~ $\Delta k\propto \nu^{-3}$ 
which is roughly consistent with our observations.
As the relativistic particles become less important relative to the cold electrons 
\citep[a criterion is given in][]{KM98} the variation tends to $\nu^{-1}$ \citep{Pach73}. 

We can consider the dominance criterion for relativistic particles 
(denoted with subscript, $er$) over thermal (`cold') particles,
(denoted, $ec$).
This criterion is ~$n_{ec}\ll\gamma_1n_{er}\ln{(2\pi\nu/\nu_g B\gamma_o)}$ (recall $\nu_g$ is the gyrofrequency, Eqn.~\ref{eqn:S_selfabs}), 
which 
requires $n_{ec}\ll  ~2.3\times 10^2$~$cm^{-3}$ if $B=0.01$, $\gamma_o\approx 5$ 
\citep[e.g.][]{BF2002} at L-band. 
This is about the same order of magnitude as the density of relativistic particles estimated in Sect.~\ref{sec:ssa} and
implies that the thermal mass in the source is probably of the same order of magnitude as the relativistic particle mass. We will pursue
this further in Appendix~\ref{appendixC}.

The only similar result of which we are aware is for the quasar PKS B2126-158 as reported in \cite{OS13}. 
They find ~$\overline{\alpha_V}$ to be $-3\pm 0.4$ at somewhat higher frequency ($8$ GHz) compared to our result. Their brightness spectrum peaks at a frequency similar to what we infer in Fig.~\ref{fig:spectra}. Moreover at lower 
frequencies, where $I_\nu\propto ~ \nu$ as in our spectrum, they detect rising circular polarization between $0.5$\% and $1$\% in what is the optically thick region. This quasar source thus displays a slightly broader transition between synchrotron self-absorption and an optically thin regime. Significantly when compared to our results, the observed linear polarization is also nearly zero near the peak of the spectrum. 
This is where their observed circular polarization is in fact largest, much as in our case. These authors explain their 
 $-3\pm 0.4$  spectral index in circular polarization in the optically thin regime, via linear to circular 
conversion  \citep{KM98} (see next subsection)
which we also adopt in part, although 
our measurements are taken closer to the optically thick transition.

\subsubsection{Conversion of Linear to Circular Polarization}

The mechanism described above, of course, requires an initial linear polarization to be rotated on the Poincar\'e sphere. 
Fortunately, \cite{BF2002} give a relatively simple set of equations together with a complete set of 
absorption and emission coefficients for a relativistic plasma, which allow us to investigate the spectral
dependence and conversion from linear to circular polarization in more detail.  We develop these in 
Appendix~\ref{appendixC} and extend their analysis to include the case $p=2$, as implied by our total intensity spectrum
(Sect.~\ref{sec:ssa}).

We consider the two limiting cases of small and large Faraday depth.  For the core of NGC~4845, 
we also show numerically that such conversion can explain the circularly polarized flux that is observed.
However, the true condition in the source is likely to be intermediate between these limits. 
The limit that is closer to what we observe is certainly the low Faraday rotation source plasma. 
 The circularly polarized flux is given by Eqn.~\ref{eq:Vconversiona}, 
as a blend of synchrotron emissivity and rotational conversion,
 which agrees with our observed ~$\overline{\alpha_V}$.  Should $B(-2)$ be as large as $10$, the required linear 
polarization for conversion is reduced to typical values ($3$ or $4$ \%). 
To reduce the linear polarization further, however (i.e. to less than 0.5\% as observed), 
it must still be suppressed by an inhomogeneous, 
depolarizing medium.  The presence of such a medium is not unusual for edge-on galaxies; a comparison to other galaxies in the 
CHANG-ES sample \citep{wie15} shows that 
a large number also lack linearly polarized flux at L-band.

We conclude that to explain our observations we need only that the relativistic particles are present and are dominant at L-band.  
The boundary conditions ~$U=U_o$ and $Q=V=0$ do suggest that we are observing an organized synchrotron source through an 
inhomogeneous, medium of mixed cold and hot particles \citep[e.g.][]{ZK2002}. For a coherent synchrotron source 
near the jet axis (see Appendix~\ref{appendixA} for the general scheme), then we may suppose that our calculations 
apply to a  sheath where relativistic particles still dominate.  The depolarizing screen around the jet possessing 
dominant Faraday rotation may gradually appear due to an influx of entrained thermal particles. 
The circular polarization incident on the screen should not be affected by the Faraday rotation in the screen. {This picture is similar to a two jet model \citep{WJ-Z14}, except that the outer jet here contains a mixture of relativistic and thermal electrons. The thermal particles presumably become dominant at larger scales perhaps due to entrainment of a surrounding thermal cloud.

\subsection{A Proposed Radio/X-ray Connection}
\label{sec:radio-X}

\subsubsection{Inverse Compton Radiation}
\label{sec:SSC}

We first consider whether inverse Compton radiation (the so-called synchrotron self-Compton, or SSC mechanism,
for up-scattering from radio to X-ray energies) can explain the X-ray emission, {\it given the properties
as measured in the radio source now}.

Inverse Compton emission is related to  synchrotron emission from an ensemble of relativistic particles through 
the ratio of the radiation energy density to the magnetic energy density,~$U_{rad}/U_{mag}$, that is,
the SSC flux would be $S_{SSC}\,=\,S_{R}\,\,U_{rad}/U_{mag}$, where $S_R$ (erg s$^{-1}$ cm$^{-2}$) is the measured radio flux.
We estimate the radiation energy density from $U_{rad}\,=\,(4\,\pi/c)\,I$, so that 
\be
\label{eqn:urad}
U_{rad}\approx \frac{16}{c}\frac{\nu S_\nu}{\theta_s^2}\approx 5\times 
10^{-7}\frac{S_{(4.87)}\nu_{(4.87)}}{\theta_s(mas)^2}~~ergs~ cm^{-3}.
\ee
We have taken the maximum from the earliest epoch in Fig.~\ref{fig:spectra} when the frequency is $4.87$ GHz
(letting $\nu_{(4.87)}=1$)
 and the flux density at that frequency
is $433$ mJy (Table~\ref{table:spectral_fits}), or $S_{(4.87)}\,=\,1$. Using~$U_{mag}=B^2/(8\pi)$, we find the ratio
\be
\label{eqn:ssc}
\frac{U_{rad}}{U_{mag}}\approx 0.12\frac{S_{(4.87)}\nu_{( 4.87)}}{\theta_s(mas)^2B(-2)^2}.
\ee
We might therefore expect inverse Compton emission during the radio outburst at about 10 percent of the radio flux density
using $B(-2)=1$ (0.01 Gauss) for the field strength (Appendix~\ref{appendixD}).
 This emission should be in a band about the peak value~$\gamma_e^2\nu_{peak}$. 
Using our previous estimate for $\gamma_e$ (estimated for L-band, Sect.~\ref{sec:circular_polarization})
 we expect this to lie in the near infrared to optical bands -- not the X-ray.

To determine the expected X-ray flux,
we can find the maximum radio flux from our observations by integrating over the band from 
0.5 to 10 GHz for
the highest curve of Fig.~\ref{fig:spectra}, finding $\approx\,2\,\times\,10^{-14}$ erg s$^{-1}$ cm$^{-2}$. 
The X-ray light curve in the 17.3-80 keV band ranges from $10^{-13}\, \rightarrow\,8\,\times\,10^{-11}$  
 erg s$^{-1}$ cm$^{-2}$ \citep[][their Fig.~8]{NW13}.  

The last point, in particular, in Fig.~8 of
\cite{NW13} overlaps in time with our observations. Whatever the origin of this X-ray emission is, it cannot be due to SSC
from the radio.
Not only is the luminosity low, but the Compton scattered frequency of the radio photons falls far below
the X-ray band  as indicated above. 
 Thus there is another source of X-ray emission independent of the SSC mechanism. In \cite{NW13} it is recognized that 
this last point does {\it not} fit the expected tidal disruption behaviour that is found near the peak X-ray emission. 
It is likely that some late behaviour arising near the accreting disc of the system 
(which is perhaps at $\sim 1$ AU from the black hole) is responsible.



\subsubsection{Inverse Compton Radiation from an Evolving Source}
\label{sec:evolving_compton}

If the radio emission does not explain the X-ray emission via SSC for current conditions, can we explain
the X-ray emission by extrapolating the radio conditions to the epoch of the peak X-ray emission?

A remarkably similar set of observations to our own were reported for the radio emission associated with a Swift 
Gamma-ray source \citep{zau11}. These authors concluded that they were seeing the birth of a parsec scale radio jet in 
the core of a rather distant galaxy.
Like us (previous subsection), these authors rejected inverse Compton emission 
because of the luminosity shortfall, and also because of the lack of correlation between the radio
and X-ray variability. 


However, in our case, the X-ray energy distribution power law near the peak emission 
 \citep[$p_X\,\approx\,2.2$,][]{NW13}
is similar to the radio energy distribution power law ($\approx\,2$, Sect~\ref{sec:ssa}) and both X-ray and radio emission
consist of a single strong point-like source at the center of the galaxy.  We are thus further motivated to attempt to
establish a potential link in the jet
 development.  A complete description of our development is given in Appendix~\ref{appendixD} and is within the context
of the jet model of Appendix~\ref{appendixA}. We summarize the arguments and results here.

Recall that timescale arguments (Sect.~\ref{sec:size}) suggest a radio source size of order $0.1$ pc ($\sim 1$ mas)
and, if the jet is relativistic (Appendix~\ref{appendixA}), then we expect that the
shocked X-ray emitting region at the earlier time (the epoch of the peak X-ray emission about 1 year earlier) is an order of 
magnitude smaller in size, a region that we refer to as the X-ray `lobe'. The observed flux density of
the circular polarization is consistent with a magnetic field strength of order $10^{-2}$ G (these are
order of magnitude estimates). Along with the observed flux density of the peak of the radio emission at time, T1,
these are the only inputs to our model, the inputs being constrained by observation. 

The extrapolation allows the magnetic field strength in the X-ray lobe to be estimated since the variation with
size is known for a constant velocity jet ($B\propto 1/r^2$, 
Appendix~\ref{appendixA}).  Since the X-ray frequency is known, an estimate of the Lorentz factor that is required
to produce the X-rays via SSC can be determined along with the critical frequency of synchrotron photons in the
X-ray source.  We find that the synchrotron emission in the X-ray lobe would peak in the infra-red.

The flux of the IR synchrotron radiation in the X-ray lobe can be found by relating the currently observed 
synchrotron emission of the radio source to the IR synchrotron emission in the X-ray lobe, knowing how
synchrotron emission depends on
the Lorentz factor, the magnetic field, the frequency, the source size, the electron energy power law ($p=2$) 
 and the electron particle constant, $N_0$,
(the latter quantity depends on time in a known fashion for an adiabatically expanding source,
see Sect.~\ref{sec:adiabatic}).  The X-ray flux from SSC, $S_{X_{SSC}}$, in the X-ray lobe can then be determined from the
IR synchrotron emission and magnetic field (previous subsection).
  
Our calculations (Appendix~\ref{appendixD}) for the above-indicated conditions, show that   
$U_{rad_{IR}}/U_{mag}\,=\,1.1$ and
$S_{{X}_{SSC}}\,=\,2.0\,\times\,10^{-11}$ erg s$^{-1}$ cm$^{-2}$, the latter value falling short of the peak observed
X-ray flux of $S_{X}\,=\,8.0\,\times\,10^{-11}$ erg s$^{-1}$ cm$^{-2}$ \citep{NW13} by a factor of 4.  
The result is, however, very sensitive to the magnetic field in the X-ray lobe.
 We have, so far, allowed
for an increase in the magnetic field to the earlier epoch that only accounts for a change in size via jet geometry,
and have not yet allowed for the possibility that shocks in the X-ray lobe may enhance the magnetic field further.
If we include a modest enhancement of the magnetic field (a factor of 1.8, whereas a strong shock would give 4),
then we find the same ratio of 
$U_{rad_{IR}}/U_{mag}$ and
$S_{{X}_{SSC}}\,=\,8\,\times\,10^{-11}$ erg s$^{-1}$ cm$^{-2}$, in agreement with the X-ray observations.

Given these results, we are encouraged to discuss such a SSC X-ray source in more detail (see Fig.~\ref{fig:cone}).
We imagine that  largely thermal material  at a very high temperature is expelled from near the black hole due to the tidal disruption of a Jupiter-sized object. 
The disrupted material also forms  a disc or torus near $1$ AU. The ejected thermal material is collimated into a fast beam by this disc. 
Subsequently the beam collides with a cloud of interstellar material at about $0.01$ pc and forms an X-ray-emitting region 
(the X-ray `lobe'). This lobe consists of a forward shock in the cloud, a reverse shock in the beam,
 and a separating contact discontinuity in the medium between the shocks. The model is much as was described in another context in \cite{HPM91}.  

For the shock wave acceleration to be effective in accelerating the electrons, the acceleration time, $t_{acc}$, must be less than the synchrotron 
lifetime. In \cite{HPM91} this was estimated in their equation (7). For relativistic shock speeds
(it is sufficient for $v_{shock}$ to be $>\,0.001c$) and our magnetic field
($B$ in the X-ray lobe of 1.8 G), then $t_{acc}\,<\,10$ min (proportionately shorter for higher $v_{shock}$).
The synchrotron lifetime of an electron (Eqn.~\ref{eq:tsync}) under these conditions 
with $\gamma_e\,=\,1109$ (Appendix~\ref{appendixD}) on the other hand, is
about 20 minutes.  Hence the acceleration time is sufficiently short. 

This result also indicates that there must be continuous injection
or acceleration of the electrons in the X-ray lobe over a time during which the SSC mechanism is producing the hard
X-ray emission.
From the X-ray light curve \citep{NW13}, this would occur around the peak of the light curve and up to $\approx\,120$ days
after the peak if the first cluster of X-ray measurements after the peak is produced in the same way. After that time,
the X-ray light curve declines precipitously (suggesting that there is no longer continuous injection/acceleration)
revealing a secondary X-ray-emitting source, possibly associated with the accretion disk.

If the reverse shock is a very strong shock, it produces an accelerated distribution of electrons with energy power law,
 $N_oE^{-3/2}$ \citep{MD2001}, i.e. the injection spectrum is described by $p_{inj}\,=\,3/2$.  
If this is the continuous injection spectrum for the electrons from this shock, then according to 
\cite{Long94}, above an energy given by
\be
E_s=\frac{1.2\times 10^7}{B_s^2t_s(1+\frac{U_{rad}}{U_{mag}})} eV
\ee
($t_s$ is the source lifetime in years and $B_s$ is the field in Gauss), the spectrum steepens from synchrotron losses such that
$p\,=\,p_{inj}+1\,=\,2.5$.
The required electrons are well above this energy and therefore we expect 
 a power law, $E^{-2.5}$, from this shock. If the forward shock is less strong, then we expect the 
test particle result, $p=2$, to be given to the electrons by this shock. 
The electrons scattered upstream of the forward shock will form the radio spectrum with $p=2$. 
These must be a significant fraction of the electrons scattering photons to X-ray energies, but not in energy space.
These accelerated electrons that leave the shock in the forward direction cool rapidly
by synchrotron radiation and eventually form the 
adiabatically expanding radio lobe that we infer from our observations. The cooling time is rapid enough to convert
the high energy electrons that leave the X-ray lobe into the low energy radio electrons that are
observed about a year after the peak.  



In summary, in the context of this model, the regions of the shocks, of size $\approx$ 1000 AU and 
magnetic field $\approx$ 1 G can produce X-ray emission via SSC  (the X-ray 'lobes') with outflow 
in the forward direction developing into the radio jet.  Relativistic electrons cool rapidly after leaving the shocked region, 
 and  develop into an adiabatically expanding outflow.

\section{Conclusions}
\label{sec:conclusions}

Our main conclusions are as follows.

1.  The radio emission from the Virgo Cluster spiral, 
NGC~4845, is dominated by a single unresolved point source that has increased by a factor of $\approx$ 6 since the 1995 NVSS
survey 
and also varied over the $\approx$ 6 month time scale of the CHANG-ES observations (Table~\ref{table:flux_densities}).
Prior to NVSS, there is no evidence for radio variability from the few data points that exist
(Sect.~\ref{sec:variability}).  However, some vertical
radio extensions as well as a distinct optical cone (Fig.~\ref{fig:optical_Cband_contours}) suggest that
previous outflow has likely occurred.

2. A small central disk of diameter,
1.8 kpc, surrounds the nucleus but constitutes only 19 mJy at L-band (less than 10\% of the total flux density)
 and  7  mJy at C-band (less than 2\% of the total).
The spectral index of this disk is $\alpha_{disk}\,=\,-0.74$ 
(Sect.~\ref{sec:flux_densities}).

 
3. The CHANG-ES observations overlap in time with the hard X-ray light curve of
 \citet[][their Fig.~8]{NW13}.  The X-ray flare, interpreted as being due to a tidal disruption event (TDE),
provides a well-defined reference point for the origin of the current radio emission.  We provide a simple
 jet model for the radio emission
(Appendix~\ref{appendixA} and Fig.~\ref{fig:cone}) which
is relevant whether the X-ray outflow originated with a TDE
or is due to infall from an accretion disk, though the former may lead to the latter in any case.

4.  Variability suggests a source size of $\theta_s \le 4 \sin{(\iota)}$ mas
(of order 0.1 pc), where $\iota$ 
is the viewing angle (Sect.~\ref{sec:size}).

5.  The wide bands used in the CHANG-ES survey have allowed us to determine in-band spectral indices and to
solve explicitly for the spectrum of the AGN as a function of time (Sect.~\ref{sec:spectra} and 
Fig.~\ref{fig:spectra}).  
At L-band, $\alpha_{1.57~GHz}\,\approx\,+1.0$ and at C-band, $\alpha_{6.00~GHz}\,\approx\,-0.5$
(Table~\ref{table:spectral_indices}), suggesting an energy spectral index for the relativistic electrons of $p\,=\,2.0$
and that NGC~4845 is a nearby example of a `GigaHertz-peaked spectrum' (GPS) radio source. 

6.  The turn-over in the spectrum can be explained by synchrotron self-absorption  (Sect.~\ref{sec:ssa}).
 However a contribution from thermal absorption at L band is possible.

7. The peak of the radio spectrum both declines and moves to lower frequencies with time (Fig.~\ref{fig:spectra}). 
This behaviour
can be understood in terms of adiabatic expansion (Sect.~\ref{sec:adiabatic}). 

8. We detect circular polarization (Sect.~\ref{sec:circular_polarization}) of order a few \% in the source, 
although no significant linear polarization, at L-band
(Table~\ref{table:flux_densities}). The circularly polarized flux has a steep spectral index
 ($\approx -3$, consistent with other authors) and can be explained by conversion from linear polarization --
 a natural consequence of generalized Faraday rotation for a source that is dominated by relativistic particles
\citep[see][and Appendix~\ref{appendixC}]{BF2002}. A further reduction of linearly polarized flux 
 requires an additional foreground
depolarizing screen.

9. Inverse Compton emission from the {\it currently observed} radio source is insufficient to account for the X-ray emission
detected by \citet{NW13}.
 
10.  However, if we extrapolate to the peak of the X-ray light curve at which time the source size is smaller
and the magnetic field higher (Fig.~\ref{fig:cone}), 
the X-ray emission can  be explained by inverse Compton upscattering by the  relativistic
particles (SSC), a result supported on energetic grounds and also by the similarity between spectral indices at radio and X-ray
wavelengths (Sect.~\ref{sec:evolving_compton}).  

\vskip 0.1truein

NGC~4845 appears to be a young nuclear radio jet/outflow and is, by far, the closest known GPS source.  As such, it 
 provides a unique opportunity to study an evolving radio source associated with a measured hard X-ray burst.
The radio source, after 5 years at an assumed
bulk outflow speed of $c/\sqrt{3}$ (Appendix~\ref{appendixA}), would have expanded to about 1 pc, or 
12 mas.  This size is within
the realm of Very Long Baseline Interferometry.
 With observations over several years, it may be possible to directly
observe the proper motion and evolution of this source, and further piece together its connection to hard X-ray outflows.

\acknowledgments

This work has been supported by a Discovery Grant to the first author by the Natural Sciences and Engineering Research
Council of Canada. 
This research has made use of the NASA/IPAC Extragalactic Database (NED) which is operated by the Jet Propulsion Laboratory, California Institute of Technology, under contract with the National Aeronautics and Space Administration. 
The National Radio Astronomy Observatory is a facility of the National Science Foundation operated 
under cooperative agreement by Associated Universities, Inc.

\newpage

\appendix

\section{The Standard Jet Model}\label{appendixA}

We are imagining a shocked section of a conical outflow \citep[or possibly the pseudo core discovered 
in][]{J88} as the source region.  Fig.~\ref{fig:cone} shows the geometry.

The jet model summarized here has been assumed as the synchrotron source in AGNs by many authors, 
beginning with \cite{BK79} and then \cite{J88}. However the dynamical details of such sources are not readily found in the literature. Moreover we are able to suggest a novel observational idea based on the net current flow through 
the jet. For these reasons we have decided to include a description of the model in this appendix. 
Ultimately it would form the basis of an analysis in the style of \cite{J88}, which is the level required to match more 
abundant observations.  The simple model described here is similar to that used by \cite{J88}. However we treat 
the magnetic field in more detail and we include a decelerating branch of the outflow. Whether this branch or the 
more familiar accelerating branch appears, depends on ambient boundary conditions.  
In addition, we consider the possibility of change in the direction of the current flow in the jet and corresponding
change in the magnetic field.

One assumes that the jet originates near the compact object as a gas dominated by relativistically hot particles. Moreover if 
$p$ is the gas pressure and $\rho$ is the total energy density (including the thermal 
energy)\footnote{We retain `standard' nomenclature that is consistent within this Appendix 
but may differ from the main
text, e.g. $p$ refers to pressure here whereas it was used as the energy spectral index in the text (likewise for
subsequent appendices).}
then we assume for simplicity 
the `polytropic' relation (we take units with $c=1$, e.g. $a_s\,c$ is the sound speed in usual units) 
\be
p=a_s^2\rho.
\ee  
One defines the conserved rest mass density $\mu$ according to ~$p+\rho=\mu(d\rho/d\mu)$,~ so that after inserting the pressure and integrating
\be
\frac{\rho}{\rho_s}=(\frac{\mu}{\mu_s})^{1+a_s^2}.
\ee
The constants with subscripts $s$ are evaluated at some convenient reference point.  

It is also useful, when expressing the dynamical equations, to define the internal energy per unit rest mass energy as 
\be
\xi\equiv \frac{p+\rho}{\mu}=(1+a_s^2)\frac{\rho_s}{\mu_s}(\frac{\mu}{\mu_s})^{a_s^2}.\label{eq:xi}
\ee

Hence also 
\be
\rho=\frac{\xi\mu}{1+a_s^2}.
\ee

We assume strictly radial flow, $u$, in the jet together with a steady state. This renders the transverse equilibrium equation useless as it only defines some unknown confining force in the theta direction. We are left with energy conservation along the flow direction together with a rest energy conservation law . These yield respectively, after a slight rearrangement, when written in the observer frame of reference  
\be
\xi\gamma(u)={\cal E},\label{eq:engcons}
\ee
and 
\be 
r^2\mu \gamma(u)v_r=A.\label{eq:masscons}
\ee
Here ${\cal E}$ is essentially the constant specific energy in the jet and $A$ is the constant rest `mass' flux through the jet.

For the energy conservation to take this simple form, there can be no Ohmic dissipation or 
other working on the jet by the electro-magnetic field. Assuming that the electric field vanishes in the 
co-moving frame (${\bf E}=-{\bf u}\wedge {\bf B}$) and that the magnetic field in the observer frame is ~${\bf B}=B_r \widehat{\bf e}_r+B_\phi\widehat{\bf e}_\phi$~ then ~${\bf E}=uB_\phi\widehat{\bf e}_\theta$~ and so we require the poloidal current 
density $j_\theta=0$ 
in order that the dissipation ~${\bf j}\cdot{\bf E}=0$. In fact one expects to have~${\bf j}\parallel {\bf B}$,~ although 
this is not consistent with purely radially moving charges. It can only be the bulk flow that moves wholly radially. 

These last two equations  in this simplified form suffice to determine the jet dynamics as a function of $r$, as we will report below. However the question as to the evolution  of the magnetic field arises. 

The rest mass conservation law and Faraday's law can be combined to give the equation satisfied by the convected magnetic field on each stream line in the form
\be
\frac{d}{dt}\left(\frac{{\bf B}}{\mu\gamma(u)}\right)=\left(\frac{{\bf B}}{\mu\gamma(u)}\cdot \nabla\right){\bf u},
\ee
which for purely radial velocity integrates to give
\bea
\frac{B_r}{\mu\gamma(u)}&=&\left(\frac{B_r}{\mu\gamma}\right)_s\frac{u(r)}{u_s}\nonumber\label{eq:mfields}\\
\frac{B_\phi}{\mu\gamma(u)}&=& \left(\frac{B_\phi}{\mu\gamma}\right)_s\frac{r}{r_s}.
\eea
Magnetic flux conservation is contained in these equations provided that ~$\nabla\cdot {\bf B}_s=0$.

To understand the significance of ~$B_{\phi s}$ we consider the radial component of the current. Restoring $c$ for 
the moment and taking account of the axial symmetry one finds at any radius\footnote{`Radius' refers to the radial direction;
the cross-sectional radius of the cone is referred to as $r\,sin\,\theta$.}  
\be
B_{\phi }=\frac{2}{cr}\frac{I_{r}(\theta)}{\sin{\theta}},\label{eq:Ir}
\ee
where $I_r(\theta)$ is the total current inside the angle $\theta$. For a zero net current through the central source, counting both the jet (polar angle $\theta_j$) and a likely sheath (polar angle $\theta_{sh}>\theta_j$), we must have $I_r(\theta_{sh})=0$. 
The only way in which the azimuthal magnetic field can reverse sign however is if there is a net current towards or away from the central source. 

For example let us assume that the radial current density through the jet is of the form
\be
j_r=\frac{I}{r^2}(\tan{\theta_+}-\tan{n\theta})\sin{\theta},
\ee
where $\tan{\theta_+}$ is a convenient parameter and $n$ is an integer that determines how narrow is the jet. Then the radial current is found to be ($n=2$)
\be
I_r=2\pi I\left(\tan{\theta_+}(\frac{\theta}{2}-\frac{\sin{2\theta}}{4})+\frac{\ln{(\cos{2\theta})}}{4}+\frac{1}{4}(1-\cos{2\theta})\right).
\ee
When $\theta_+=\pi/6$ and $n=2$ the current passes through zero at $\theta=19.7^\circ$. Beyond this angle the net current is negative.  A balancing current must flow in the more diffuse surroundings or on the surface in order to prevent the accumulation of central charge. 
Such shearing magnetic fields (e.g. \ref{eq:Ir}) can lead to circular polarization \citep[e.g.][]{ZK2002} and possible rapid variations in the observed rotation measure (George Heald-private communication).

The interaction of the radial current density and the azimuthal magnetic field leads to a `pinching' or collimating force. The parameters might be chosen so as to balance the transverse equation at the reference point or `nozzle'. However the assumed conical geometry is not generally consistent with the subsequent transverse force equation, so we do not attempt this here.

The model does not yet allow for the acceleration of the jet by the magnetic field. The Poynting flux in the 
radial direction due to the wound-up `spring' component of the field at any point $r_s$ is ~$u(r_s)B_{\phi s}^2/4\pi$. The associated energy density is then ~$B_{\phi s}^2/4\pi$~. One might try to include this effect by adding to $a_s^2$ in the expression for pressure, the quantity ~$a_a^2\equiv B_{\phi s}^2/(4\pi\rho_sc^2)$. However we shall find that this quantity is not constant in radius, so the approximation can only apply over a small section of the jet compared to its length.
We will also ignore this possibility in what follows, but it can be restored locally simply by modifying $a_s^2$ to $a_s^2+a_a^2$.

The jet velocity profile follows from Eqn.~\ref{eq:engcons} and Eqn.~\ref{eq:masscons}, 
once $\mu$ is eliminated in favour of $\xi$ and hence $\gamma$. We can eliminate most of the parameters by writing ~$x=r/R$,~ where (after restoring $c$) 
\be
R^2\equiv \frac{A}{\mu_sc}\left(\frac{\rho_s(1+a_s^2)}{\mu_s {\cal E}}\right)^{1/a_s^2}.\label{eq:R2}
\ee
 The velocity in the jet as a function of scaled radius in units of $c$ is then found from
\be
x^2=\frac{1}{\beta(1-\beta^2)^{\frac{1-a_s^2}{2a_s^2}}}.\label{eq:x}
\ee 

This expression gives the velocity at $x$ for any particular value of $a_s^2$ although it contains some surprises. As $x\rightarrow\infty$ one can have two branches for $a_s^2<1$. In the accelerated branch $\beta\rightarrow 1$, while in the decelerated branch ~$\beta\rightarrow 0$. The choice is made by the boundary conditions at infinity, which apply a finite pressure in the decelerated jet flow and vanishing pressure in the accelerated jet flow. 

It is also clear that there is a minimum radius at which these two branches can exist. If we calculate ~$dx^2/d\beta=0$ we find that at this minmum, which we may take to be $x_s$, 
\bea
\beta&=&a_s,\nonumber\\
x_s^2&=&\frac{1}{a_s(1-a_s^2)^{\frac{1-a_s^2}{2a_s^2}}}
\eea 
The minimum point represents therefore the origin of the jet in a `sonic' launch point. The two branches meet at this point and there is no flow extending to smaller radius with $\beta>0$. This behaviour is due to the absence of a restraining force such as gravity. The presence of gravity would turn this minimum point into a true choke point in the jet flow or wind. Once 
Eqn.~\ref{eq:x} is solved we find $\xi$ from Eqn.~\ref{eq:engcons} and hence $\mu$, $\rho$, $B_r$ and $B_\phi$ from the various relations. 

We can be more explicit in the interesting case where $a_s^2=1/3$ (that is, the jet velocity is
the sound speed expected for relativistic internal energy, $u\,=c/\sqrt{3}$, when $c$ is restored).
 Eqn.~\ref{eq:x} becomes the parameter-free cubic equation
\be
\beta^3-\beta+\frac{1}{x^2}=0,\label{eq:cubic}
\ee
and $x_s^2=\sqrt{27}/2$ and $\beta_s=1/\sqrt{3}$. The discriminant of this equation is 
$\Delta=-4+27/x^4\equiv 4(-1+x_s^4/x^4)$. When $\Delta>0$ so that $x<x_s$ there is only one real root and it is negative. At $x=x_s$ we have $\Delta=0$ and there is only one real (double) root $\beta_s=1/\sqrt{3}$. 

In the region of interest, $x>x_s$, we have $\Delta<0$ and there are three real roots but only two are positive. The positive roots take a convenient form; for the accelerated branch as 
\be
\beta_{acc}=\frac{2}{\sqrt{3}}\cos{\left(\frac{1}{3}\arccos{(-\frac{\sqrt{27}}{2x^2})}\right)},\label{eq:jacc}
\ee 
and for the decelerated branch as
\be
\beta_{dec}=\frac{2}{\sqrt{3}}\cos{\left(\frac{4\pi}{3}+\frac{1}{3}\arccos{(-\frac{\sqrt{27}}{2x^2}})\right)}.\label{eq:jdec}
\ee

These give respectively $\beta_{acc}\rightarrow 1$  and $\beta_{dec}\rightarrow 0$ as $x\rightarrow\infty$. At $x=x_s$ each expression yields the root $\beta_s=1/\sqrt{3}$ as expected. The general behaviour is illustrated in 
Fig.~\ref{fig:jets}. The curves actually meet vertically at $x_s$. 

\begin{figure*}[!ht]
   \centering
   \includegraphics*[width=0.5\textwidth]{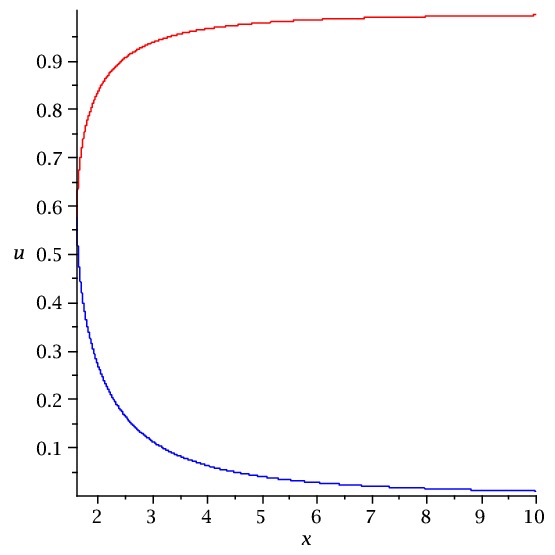}
   \hspace{-1.20in}
   \caption{The upper curve is the accelerated jet universal velocity behaviour when $a_s^2=1/3$ in terms of $x=r/R$. The lower curve is the decelerated jet velocity under the same conditions. The pressure at infinity goes to zero to establish the upper curve, while it retains a finite value on the lower curve. the curves meet vertically at the central value of $u=1/\sqrt{3}=0.57735$.}
\label{fig:jets}
\end{figure*}

The behaviour at large $x$ is easy to determine using Eqn.~\ref{eq:jacc} and Eqn.~\ref{eq:jdec}. For the accelerated jet one finds in lowest order ($\asymp$ means equals in the asymptotic limit)
 
\bea
\beta&\asymp& 1-\frac{1}{2x^2},\nonumber\\
\gamma&\asymp& x,\nonumber\\
\xi&\asymp& \frac{{\cal E}}{x},\label{eq:jaccasymp}\\
\mu&=& (\frac{3\mu_s}{4\rho_s})^3\xi^3,\nonumber\\
\rho&=& \frac{3}{4}\xi\mu.\nonumber
\eea
Eqns.~\ref{eq:mfields} then show that asymptotically $B_r\propto x^{-2}$ and $B_\phi\propto x^{-1}$ as is normally assumed. 
 The behaviour of the magnetic field near the source is however more complicated. This is the branch 
describing the jet expanding into a low pressure ambient medium.

For the decelerated jet the behaviour follows from Eqn.~\ref{eq:jdec} as
\bea
\beta&\asymp& \frac{1}{x^2},\nonumber\\
\gamma&\asymp& \frac{1}{\sqrt{1-1/x^4}},\label{eq:jdecasymp}\\
 \xi&=& {\cal E}\sqrt{1-1/x^4},\nonumber
\eea
and the expressions for $\mu$ and $\rho$ follow from the dependences on $\xi$ given in Eqns.\ref{eq:jaccasymp}). 
Boundary conditions at infinity are determined by condtions at the origin $x_s$ and vice versa. The radial field varies again as $x^{-2}$ in lowest order, but the azimuthal field increases as ~$B_\phi\propto (x-1/x^3)$. Even in the absence of shocks, this can lead to an outer brightening of the jet. However a dense medium at infinity such that ~$\xi_\infty={\cal E}$~ is required to decelerate the jet. This dense medium may be comprised of material swept-up from the ambient medium. 

\newpage

\section{Source Proper Frame Properties}
\label{appendixB}

We have used quantities in the local rest frame for our calculations. This ignores the 
possible relativistic motion of the source. In the presence of rest frame gas against 
which the source is expanding, a deceleration to non-relativistic source velocity is possible (see e.g. 
Appendix~\ref{appendixA}). 
However it is important to be aware of the modifications required by relativistic source motion. 
We discuss these in this section. Since NGC 4845 is not at large cosmological redshift, we do not include this effect.

In the first instance we have argued generally in terms of the standard isotropic, energy power-law, distribution 
of source electrons (or possibly positrons).
However a particle moving in the source frame at an angle $\theta'$ to the relative velocity, appears at 
an angle $\theta$ to the velocity in the local rest frame where $\theta$ is (e.g. \cite{H11})
\be
\tan{\theta}=\frac{\sin{\theta'}}{\gamma_u(u)(\cos{\theta'}+u/v')},
\ee
where $u$ is the relative velocity,~ $v'$ is the source particle speed, and $\gamma_u$ is the bulk flow Lorentz factor. 
The inverse relation is found by interchanging the primes  and replacing $u$ by $-u$.  

This transformation shows that only with $u/v'\ll 1$ and $\gamma_u\approx 1$ do the angles coincide and hence that isotropy is invariant. Otherwise, even a source particle moving at  $\theta'=90^\circ$ when $\gamma_u=3$ ($\beta\approx 0.94$) and $u/v\approx 1$ is directed at ~$\theta \approx 18.4^\circ$ in the local rest frame. This cone contains all other forward moving particles. Particles with $\theta'>90^\circ$ are also beamed to positive angle in the rest frame.

Consequently, an isotropic distribution of particles in the source frame is a directed beam in the observer frame. 
Fortunately the beaming due to synchrotron radiation itself tends to dominate this effect in the integration over solid 
angle. 

The energy of the particles transforms according to 
\be
E=\gamma_u(1+\frac{{\bf u}\cdot{\bf v}'}{c^2}) E',~~or~~ \gamma=\gamma_u(1+\frac{{\bf u}\cdot{\bf v}'}{c^2})\gamma',
\ee
where ${\bf v}'$ is the source particle velocity. The inverse follows by interchanging the primes and reversing the sign of ${\bf u}$.  We see that the energy of forwardly moving particles is increased by the roughly constant factor ~$2\gamma_u$ . This implies that the power law distribution in energy in the source frame changes only its limits in the rest frame. 

It is of interest  to clarify the transformation of intensity or `brightness', $I_\nu$. this was discussed originally in this context in the paper \cite{BK79}, but it is worth repeating the details. 

The radiation pattern of a source transforms as (e.g. \cite{H11})
\be
\frac{d^2E}{d\Omega dt}=\frac{1}{\gamma_u^4\kappa^3}\frac{d^2E'}{d\Omega'dt'},
\ee 
where ~$\kappa\equiv (1-\widehat{\bf k}\cdot \frac{{\bf u}}{c})$ and $\Omega$ denotes solid angle. The unit vector $\widehat{\bf k}$ lies along the direction of a ray. For a distant observer however the time is $t_{obs}=t+R/c$ and hence ~$dt_{obs}=dt\kappa$, given $R$ as the distance to the observer along the emitted ray. Consequently we have 
\be
\frac{d^2E}{d\Omega dt_{obs}}=\frac{1}{\gamma_u^4\kappa^4}\frac{d^2E'}{d\Omega'dt'}.
\ee
However to calculate the relation between the emissions per unit frequency we must use the D\"oppler shift relation ~$d\nu=d\nu'/(\gamma\kappa)$ to obtain  

\be
\frac{d^2E}{d\Omega dt_{obs}d\nu}=\frac{1}{\gamma_u^3\kappa^3}\frac{d^2E'}{d\Omega'dt'd\nu'}.
\ee
Finally, to obtain the transformation of the intensity, we must consider an area transformation. It is not simply given in terms of the Lorentz contraction since the information about the moving surface is delivered by the light rays crossing it, rather than from measurements made by two different but simultaneous observers. If we imagine a surface element $d\sigma'$ oriented perpendicular to the plane of $\widehat{\bf k}$ and ${\bf u}$ with its normal direction at an angle $\theta'$ to the relative velocity (hence parallel to a ray in that direction), then the component~ $d\sigma'\sin{\theta'}$~ is affected by the change in reference frame. This component becomes equal to $d\sigma\sin{\theta}$ in the reference frame after tracing the light rays (which define in fact the angle transformation) . Hence we have by equating these two expressions that 
\be
\frac{d\sigma'}{d\sigma}=\frac{\sin{\theta}}{\sin{\theta'}}=\gamma_u\kappa,
\ee
where we used the standard angle transformation for light rays to get the last expression on the right. In the special case of longitudinal motion so that $\theta'=\theta=0$, this last expression continues to hold (see e.g. \cite{EW2000}).

We are now in possession of the transformation of intensity in the form

\be
I_\nu(\nu)\equiv\frac{d^2E}{d\Omega dt_{obs}d\nu d\sigma}=\frac{1}{\gamma_u^2\kappa^2}\frac{d^2E'}{d\Omega'dt'd\nu'd\sigma'}\equiv \frac{I_{\nu'}(\nu')}{\gamma_u^2\kappa^2},
\ee
which must be used together with ~$\nu'=\nu \gamma_u\kappa$. The inverse may be found in the usual way. We see that the spectral index is essentially invariant between the two frames over a small range in angle, which is expected for synchrotron radiation. The comoving radiation energy density is  reduced from that measured in the rest frame. The co-moving frequency is reduced by the D\"oppler factor which might be thought to encourage self-absorption in the co-moving frame. However the electron number density is also reduced by $\gamma_u$. If moreover the electric field in the source frame is zero, then the perpendicular magnetic field transformation is given by ~$B'_\perp=B_\perp/\gamma_u$. This reduction also offsets the frequency reduction. In the end the source absorption coefficient is virtually invariant (for $p =2$) relative to the rest coefficient.

Another important invariant is the polarization of the source, which we discuss in Sect.~\ref{sec:circular_polarization} and Appendix C. 
This invariance is due to the invariance of the phase of the electromagnetic wave.

In summary, the above relations show that, unless the jet outflow is highly relativistic,
i.e. $\gamma_u\,>>\,1$, measurements in the observer's rest frame are adequate to describe the instrinsic angles, energy 
and magnetic fields of the source.  Note that $\gamma_u\,\sim\,1$ still encompasses the possibility of very rapid jet
speeds.  For example, $u\,=\,0.5\,c$ $=>$ $\gamma_u\,=\,1.15$.

We have neglected the cosmological red-shift since NGC4845 is so close. However, this affects the observed 
frequency by a well known conversion from the local rest frame frequency ($\propto 1/(1+z)$).
\newpage

\section {Spectral Evolution of a Simple Ensemble of Relativistic Electrons}
\label{sec:simple}

Let us initially assume that the source is, in part, a stationary ensemble of relativistic electrons 
that is the result of a unique acceleration and that the event that has produced the acceleration occurred at
the time of the hard X-ray flare at time, $t_0$.

For an individual electron, the synchrotron lifetime, $t_s$, may be written as 
\be
t_s=\frac{464}{B_\perp(-2)^{3/2}\nu_5^{1/2}}~~ years,\label{eq:tsync}
\ee
where $B_\perp(-2)$ is the perpendicular (to the electron orbit) magnetic field in units of $10^{-2}$ G 
and $\nu_5$ is the peak frequency in units of $5$ GHz. The critical frequency, 
essentially the peak frequency produced by electrons of a given energy, implies that 
\be
\gamma_e=345\frac{\nu_5^{1/2}}{B_\perp(-2)^{1/2}}.\label{eq:gammaelec}
\ee
where $\gamma_e$ is the Lorentz factor of the relativistic electron radiating at 5 GHz.

For an ensemble of particles in a magnetic field, emission near the spectral turnover 
will be dominated by those electrons
of energy, $\gamma_e\,m_e\,c^2$, where $m_e$ is the electron mass and $c$ the speed of light.  Thus, from Eqn.~\ref{eq:tsync},
one expects the peak frequency to decline as $(t/t_0)^{-2}$. 

From Table~\ref{table:spectral_fits}, we see that the peak frequencies are in the ratio 
$\nu_{T2}/\nu_{T1}\,=\,4.03/4.87=0.83$ after $T1\,+\,56$ days, and in the ratio 
$\nu_{T3}/\nu_{T1}\,=\,3.21/4.87=0.66$ after $T1\,+\,196$ days. Taking $T1$ to be 342 days after the X-ray peak at
$t_0$ (Sect.~\ref{sec:discussion}),
the expected ratios are $((T1+56)/T1)^{-2} = 0.74$ and $((T1+196)/T1)^{-2} = 0.40$, respectively.
The dependence of the peak 
frequency on time is much closer to being linear than 
it is to being the quadratic variation expected from such a stationary ensemble emitting near the peak of the spectrum.

For $\nu_5\,=\,1$ (a spectral turnover of $5$ GHz, Fig.~\ref{fig:spectra}),
for fields even as large as $0.04$ G 
\citep[a comfortable upper limit, see][]{ACT08} 
the lifetime of particles radiating at the spectral turnover is about $58$ 
years and $\gamma_e=173$.
 Higher energy particles by about a factor of one hundred would be evolving substantially. 
However the corresponding frequency is in the infrared.   
 Allowing for the magnetic field to change over the course of the observations is the basis of our explanation of the frequency peak time dependence in the SSA model (see Sect.~\ref{sec:adiabatic}). 
However in this case either the $t^{-1}$ or $t^{-2}$ time dependence in magnetic field gives quite the wrong time dependence for the peak frequency. Moreover the evolution timescale becomes longer with a weakening field which results in an even poorer fit. 
We observe substantial evolution 
in less than half a year at GHz frequencies, however, which tends to exclude this simple picture even if the peak frequency ratios were acceptable.

We conclude that the behaviour of a source in which there has been a unique acceleration of 
electrons and whose emission is now fading with time does not match the observations.

\newpage

\section{Thermal Absorption}
\label{sec:thermal_absorption}

Can thermal absorption explain the low frequency cutoff?
 The thermal absorption coefficient may be written, by ignoring some small logarithmic terms, as \citep{Pach70}
\be\label{eqn:kappa_thermal}
\kappa_{{ec}_\nu} \approx \frac{n_{ec}^2}{T_4^{3/2}\nu(9)^2}\times 10^{-25} ~cm^{-1},
\ee
where the subscript, $ec$, refers to `cold' (i.e. non-relativistic, or thermal) electrons,
 the frequency, $\nu(9)$, is in GHz and the temperature is in units of $10^4$~K. Multiplying this by 
the line of sight distance, $s$, and requiring the
 product to be one (for unity optical depth) shows that, on a scale of $10^{17}$ cm in the optically thick
limit ($\nu(9)\,\sim\,1$),
 we need a cold 
 plasma density of $n_{ec}\approx 10^4~ cm^{-3}$. 
This might be in a `sheath' around the source (i.e. within or very near the cone).
On a scale of one kpc one needs only $\approx 60~ cm^{-3}$. Such an `HII region' might easily 
intervene along the line of sight in an edge-on galaxy.

However if we calculate the effective spectral index, $\alpha_{eff}$, i.e. the observed spectral index that results from
a synchrotron spectrum, $I_\nu(o)=C \nu^\alpha$, with an additional absorbing thermal screen, then the observed
spectrum will be
\be
\label{eqn:thermalabsorption}
I_\nu\,=\,I_\nu(o)e^{-(\kappa_{{ec}_\nu} s)}\,=\,C \nu^\alpha e^{-(\kappa_{{ec}_\nu} s)}.
\ee
Differentiating this spectrum with respect to frequency
and noting the $\nu^{-2}$ dependence of $\kappa_{{ec}_\nu}$ (Eqn.~\ref{eqn:kappa_thermal})
results in an observed spectral index of
\be
\alpha_{eff}=\alpha+2\kappa_\nu s.
\ee

For a non-thermal spectral index of $\alpha=-0.5$ (Table~\ref{table:spectral_indices}),
then we calculate for $\kappa_{ec_\nu} s=1$ that $\alpha_{eff}=1.5$, which is too steep 
(again, Table~\ref{table:spectral_indices}).
At $\nu(9)\ge 3$, the absorption coefficient is down by a factor ten. Therefore we obtain $\alpha_{eff}=-0.3$, 
which does not fit the spectra in Fig.~\ref{fig:spectra}. 
Therefore, thermal absorption does not describe the turn-over in the spectrum and
synchrotron self-absorption (Sect.~\ref{sec:ssa}) is preferred.  A more detailed analysis based on best fits to the data is given in the text, but our basic conclusion remains.

\newpage

\section{Circular and Linear Polarization in the Beckert \& Falcke Formulation}
\label{appendixC}

A detailed calculation of the circularly polarized flux from the jet model
described in Appendix~\ref{appendixA} must be left for elsewhere. This calculation has been done in \cite{J88}, 
although frequency dependences for the transition region were not given in that paper.  However, a quantitative 
explanation of how linear polarization is converted to circular polarization is indeed possible,
as will be described here. 

In this section,
we adhere closely to the formulation in the appendices of \cite{BF2002}. 
We adopt an isotropic relativistic plasma with a power law energy distribution having $p=2$ ($s$ in their notation),
as suggested by our spectral index in the optically thin limit (Table~\ref{table:spectral_indices}).
This plasma state implies that ~$\kappa_U=\eta_U=h_Q=0$~, using the notation of \cite{BF2002} for the U 
absorption coefficient, the U emissivity, and the extraordinary wave conversion coefficient. Our only 
additional assumptions are to take ~$\kappa_V$ as negligible and to neglect $\kappa_Q^2V$ compared to $d^2V/ds^2$. The former assumption is justified by the  stonger 
decline with frequency (by a factor $\sqrt{\nu_{B_\perp}/\nu}$) of $\kappa_V$  compared to the remaining absorption coefficients. The latter assumption neglects the second order change in $V$ by absorption over the whole source, compared to a local second order change. 

The ultrarelativistic conversion coefficient given in \cite{Saz69} requires
 taking a limit as $p\rightarrow 2$ which we obtain as 
\be
\kappa_c=2\kappa_o(\nu)(\frac{\nu_{B}}{\nu})(\sin{\phi})^2\ln{\left(\gamma_o\sqrt{\frac{\nu_B}{\nu}}\right)},
\ee 
where the scaling opacity (proportional to the Faraday absorption coefficient $\kappa_F$) is
\be
\kappa_o\equiv \frac{\pi \nu_B}{c}\frac{\nu_p^2}{\nu^2}.
\ee
The plasma frequency $\nu_p$ is 
\be
\nu_p^2=\frac{n_e e^2}{\pi m_e},
\ee
and $\gamma_o$ is the minimum Lorentz factor in the distribution. 

A straightforward but tedious manipulation of the equations in 
Appendix A of \cite{BF2002} using these assumptions, yields the following solution for the Stokes parameters as a function of path length
 in the source, $s$:
\bea
V_\nu&=&\frac{\eta_V}{\kappa_I\kappa^2}(\kappa_I^2-\kappa_Q^2+\kappa_F^2)(1-e^{-\kappa_Is})+\frac{\eta_V\kappa_I}{\kappa^4}(\kappa_c^2-\kappa_I^2)e^{-\kappa_Is}(1-\cos{\kappa s}+(\tan{\Phi})\sin{\kappa s}),\nonumber\\
U_\nu&=&\frac{\eta_V}{\kappa_c\kappa^3}(\kappa_c^2-\kappa_I^2)(\kappa_Ie^{-\kappa_Is}(\sin{\kappa s}+(\tan{\Phi})\cos{\kappa s})-\kappa),\\
Q_\nu&=& \frac{\kappa_c}{\kappa_F}V+\frac{\eta_V\kappa_I}{\kappa_c\kappa_F\kappa^2}(\kappa_c^2-\kappa_I^2)(e^{-\kappa_I s}(\cos{\kappa s}-(\tan{\Phi})\sin{\kappa s})-1).\nonumber\label{eq:Stokes}
\eea
The total intensity may be found from
\be
I_\nu=\frac{\eta_I}{\kappa_I}(1-e^{-\kappa_Is})-\frac{\kappa_c\kappa_Q}{\kappa_F}e^{-\kappa_Is}\int^s~V_\nu~ds+\frac{\kappa_Q}{\kappa_F}(U_oe^{-\kappa_Is}-U_\nu)+U_oe^{-\kappa_Is}.\label{eq:StokesI}
\ee
We note that in an optically thick region ($\kappa_IL$ large and $\kappa_Q/\kappa_F$ small), $I_\nu\approx \eta_I/\kappa_I$, 
which is proportional to ~$\nu^{5/2}$ as expected (in this appendix, $L$ is the total path length through the source, whereas $s$ is
a general path length coordinate).

 The initial conditions are ~$V(0)=Q(0)=0,~I(0)= U(0)=U_o$ and we have set to zero an arbitrary constant of integration that gives a spurious linear increase of $V$ with $s$. The coefficients, $\kappa_I$, $\kappa_Q=(9/13)\kappa_I$, $\kappa_F$, and the emissivity,
 $\eta_V$, are as in \cite{BF2002}. In addition we have defined
\be
\kappa^2=(\kappa_c^2-\kappa_Q^2+\kappa_F^2),
\ee
and 
\be
\tan{\Phi}=\frac{\kappa}{\kappa_I}+\frac{\kappa_c\kappa^3}{\kappa_I(\kappa_c^2-\kappa_I^2)}\frac{U_o}{\eta_V}.
\ee
  
The linear dependence on $\eta_V$ is a result of isotropy, our boundary conditions and the identity ~$\kappa_I\eta_Q-\kappa_Q\eta_I=0$~, which hold for the coefficients \citep{BF2002}. However the solution in detail as displayed below shows that $V$ is mainly dependent on $U_o$, as is expected. 

We turn to an analysis of the frequency dependences of these solutions. It is convenient to express all frequencies in terms of one GHz (denoted $\nu_9$), and to let all emission and absorption coefficients have their frequency independent values at one GHz when expressed with their usual names~$()$ modified to~$\tilde{()}$ . Thus~ $\eta_V=\tilde\eta_V/\nu_9$~ and~ $\kappa_c=\tilde\kappa_c/\nu_9^3$, ~$\kappa_I=\tilde\kappa_I/\nu_9^3$~ and so on. Only $\kappa$ requires a slight complication since ~$\kappa_F=\tilde\kappa_F/\nu_9^2$. Hence from the definition, $\kappa$ may either have the form $\kappa=\tilde\kappa/\nu_9^3$ if $\kappa_c$,~$\kappa_Q$ 
are dominant, or alternately $\kappa=\tilde\kappa_F/\nu_9^2$~ if $\kappa_F$ is dominant. Other than in these extreme cases, ~$\kappa$~ presents a mixed frequency behaviour.

In the next two subsections, we proceed by assuming that $\kappa_IL\ll1$ ($L$ is the path length through the source) and consider
the limits of small and large Faraday depths.

\subsection{The Case of Small Faraday Depth}
\label{sec:small_faraday}

In the first case, let us assume that ~$\kappa_F$~ is small compared to ~$\kappa_c$~or $\kappa_Q$. 
we note that this requires that the minimum Lorentz factor $\gamma_o$ be essentially equal to the $\gamma_e$ associated with the peak frequency in the radio band. Then the solution for $V$ becomes, to first order in $\kappa_IL$ and $\kappa L$, 
\be
V\approx \eta_VL\frac{\kappa_I(\kappa_c^2-\kappa_I^2)}{\kappa^3}(\tan{\Phi})+\eta_VL(1+\frac{\kappa_I^2-\kappa_c^2}{\kappa^2}).
\ee

 In addition taking $\kappa_F$ small, the expression for the circularly polarized intensity (cpi) becomes with the frequency dependences displayed
\be
V\approx \tilde\eta_VL\frac{\tilde\kappa_I(\tilde\kappa_c^2-\tilde\kappa_I^2)}{(\tilde\kappa_c^2-\tilde\kappa_Q^2)^{3/2}}\frac{\tan{\Phi}}{\nu_9}+\tilde\eta_VL\frac{\tilde\kappa_I^2-\tilde\kappa_Q^2}{\tilde\kappa_c^2-\tilde\kappa_Q^2}\frac{1}{\nu_9},
\ee

where 
\be
\tan{\Phi}=\frac{\sqrt{\tilde\kappa_c^2-\tilde\kappa_Q^2}}{\tilde\kappa_I}+\frac{U_o}{\tilde\eta_V}\frac{\tilde\kappa_c}{\tilde\kappa_I}~\frac{(\tilde\kappa_c^2-\tilde\kappa_Q^2)^{3/2}}{\tilde\kappa_c^2-\tilde\kappa_I^2}~\frac{1}{\nu_9^2}.
\ee

Consequently after inserting this value we have simply 
\be
V\approx \frac{\tilde\eta_VL}{\nu_9}+\frac{U_o\tilde\kappa_cL}{\nu_9^3}.\label{eq:Vconversiona}\label{eq:Va}
\ee

Thus when the Faraday absorption is small compared to  relativistic conversion, there is an internal 
generation term in the circular polarized flux ~$\propto \nu^{-1}$~ plus a conversion term  ~$\propto \nu^{-3}$.  
It is the
latter term that we are particularly interested in since this is the term that dicates the linear to circular
polarization and also has the steep spectral behaviour that is 
observed (Table~\ref{table:spectral_indices}). However, the first term could also contribute since the
 spectral indices, on average,
tend to be slightly flatter than -3.

A numerical calculation using the values given in \cite{BF2002} and our formula for 
$\kappa_c$ gives ($L(17)$ is in units of $10^{17}$ cm and $\gamma_o=200$)
\bea
\tilde\kappa_cL&\approx& 7.7\times 10^{-3}L(17){B^2(-2)n_e(100)},\nonumber\\
\tilde\eta_VL&\approx& 0.28B^2(-2)n_e(100)L(17)\times 10^8\,\{=\,22\,\theta_s^2\,B^2(-2)n_e(100)L(17)~~mJy\}.\label{eq:etaVnum}
\eea
The actual circularly polarized flux measured at the antenna ~$V_{\nu A}=\Omega_s V$.
Thus the first term in Eqn.~\ref{eq:Va} must be multiplied by the factor $\Omega_s$,
This factor is ~$\approx 7.8\times 10^{-7}\theta_s^2$, so that the numerical factor in the second of Eqns.~\ref{eq:etaVnum}
 becomes ~$22~~\theta_s^2$. The source angular size is expressed in milli-arcsec, and may be as small as $0.1$ when $L(17)=1$.
 For the second term, we will consider
$U_o$ as if it has already been corrected to flux density, thus Eqns.~\ref{eq:Vconversiona}
and \ref{eq:etaVnum} together give

\be
V\approx \frac{22\,{\theta_s}^2\,B^2(-2)n_e(100)L(17)}{\nu_9}+
\frac{7.7\times 10^{-3} U_o  L(17)B^2(-2)n_e(100)}{\nu_9^3}~~mJy.
\ee

Hence the first term in $V$ may contribute to our circular polarization if ~$B(-2)\approx 4$~ 
and the relativistic electron density ~$n_e$ ~ is as large as $100 cm^{-3}$ at 1.5 GHz ($V\,=\,2.3$ mJy).
Under these conditions the second term would require a linearly polarized flux $U_o\approx 60$ mJy, in order to 
produce the observed circularly polarized flux by conversion ($V\,=\,2.2$ mJy), i.e. comparable values that are roughly consistent with
our total circularly polarized flux density.
It is the second term that is favoured by our measured in-band values of ~$\overline{\alpha_V}$ however. 
The required incident $U_o$ that makes the second term comparable to the first term is larger than any linearly polarized flux (close to $24$ \%) that we or others observe. We must then consider the exit values of $U$ and $Q$ further below, especially after passing through a possible depolarizing screen \citep[cf.][]{OS13}.

The dominant term in the total intensity is approximately the first term in  Eqn.~\ref{eq:StokesI}. Assuming that 
both terms in Eqn.~\ref{eq:Va} are comparable, we obtain a rough estimate of the percentage cpf as
\be
\frac{\tilde\eta_V}{\tilde\eta_I(\nu_9)^{1/2}}\approx -\cot{\phi}\sqrt{\frac{B_\perp(-2)}{\nu_9}}\times 0.9\times 10^{-2}.
\ee
This can be of order $2$ \% for $B(-2)\ge 4$. 

 We note that any balanced mixture of the two terms in Eqn.~\ref{eq:Va}  will change quickly with increasing frequency towards a $\nu^{-1}$ behaviour. Moreover the cpi can not be much larger than $\tilde\kappa_c L U_o$
\citep[cf.][]{OS13}.  
Because of the harmonic functions of ~$\kappa s$~ in the general expression for $V$, this cpi state also changes rapidly in the optical depth transition zone as we require.
 
Unfortunately, proceeding in the same fashion as for the circular polarization, we find that the Stokes U is modified only by absorption. It becomes 
\be
U=U_o(1-\frac{\tilde\kappa_I L}{\nu_9^3}),
\ee
and so is essentially unchanged from $U_o$. The latter quantity is thus to be constrained by the small value of linearly polarized flux (lpf) that we observe. One must  appeal to a depolarizing screen to remove the linear polarization \citep[e.g.][]{OS13} if it is really of this order. We note  that a magnetic field as large as ~$B(-2)=10$ \citep[e.g.][]{ACT08}
 would reduce the required $U_o$ to a more reasonable $3-4$\%, but this would still have to be removed by an inhomogeneous, depolarizing screen to satisfy our limit.

Finally we calculate Stokes Q in the same approximation and limit. When there is very little Faraday rotation we find that ~$Q\approx 0$ in lowest order provided that ~$\tilde\kappa\approx \tilde\kappa_c$~, and that this quantity is much larger than either $\tilde\kappa_I$ or $\tilde\kappa_Q$.

\subsection{The Case of Large Faraday Depth}
\label{sec:large_faraday}

A second limiting case arises when the Faraday depth is large so that $\kappa_F$ is large compared to $\kappa_c$ or $\kappa_Q$ ~(and therefore $\kappa_I$ since for $p=2$ this coefficient is larger only by the factor $13/9$) . 
In this limit ~$\kappa\approx\tilde\kappa_F/\nu_9^2$ 

and we find  at small $\kappa_IL$ that 
\be
 V\approx \frac{\tilde\eta_VL}{\nu_9}+\tilde\eta_V\frac{\tilde\kappa_I(\tilde\kappa_c^2-\tilde\kappa_I^2)}{\tilde\kappa_F^4}~\frac{\tan{\Phi}}{\nu_9^4}\sin{(\kappa_FL)}.
\ee
Moreover 
\be
\tan{\Phi}=\frac{\tilde\kappa_F}{\tilde\kappa_I}~\nu_9+\frac{\tilde\kappa_c}{\tilde\kappa_I}~\frac{\tilde\kappa_F^3}{\tilde\kappa_c^2-\tilde\kappa_I^2}~\frac{U_o}{\tilde\eta_V}~\nu_9,
\ee
and hence
\be
V\approx \frac{\tilde\eta_V}{\nu_9^3}\frac{\tilde\kappa_c^2-\tilde\kappa_I^2}{\tilde\kappa_F^3}\sin{(\kappa_FL)}+\frac{\tilde\eta_VL}{\nu_9}+\frac{U_o\tilde\kappa_c}{\tilde\kappa_F\nu_9^3}\sin{(\kappa_FL)}.\label{eq:Vconversionb}
\ee
The terms in ~$\tilde\eta_V$~ must be assumed to be reduced numerically by $\Omega_s$ to give the observed flux, as in the previous case.

The first term in Eqn.~\ref{eq:Vconversionb} is small compared to the second term around $\nu_9=1$ according to our limiting assumption. Thus the combined frequency dependence due to the second and third terms is not changed from the small Faraday rotation limit. The magnitude of the conversion term (proportional to $U_o$) is however reduced by the factor ~$\tilde\kappa_c/\tilde\kappa_F$, which is assumed to be quite small. In fact this ratio is \citep[see formulae in][]{BF2002}
\be
\frac{\tilde\kappa_c}{\tilde\kappa_F}\approx 0.75\times \frac{\nu_B\gamma_o^3}{\nu}~~\frac{\sin^2{\phi}}{\cos{\phi}}(1+\frac{\ln{\nu_B/\nu}}{2\ln{\gamma_o}}).
\ee
This is very small unless $\gamma_o$ approaches $100$ or more, which becomes the previous case. Hence in this case the conversion term is not important compared to the synchrotron emission term given in Eqn.~\ref{eq:etaVnum}. This can give sufficient cpf as we have seen, 
and it is not oscillatory. However the frequency dependence is $\nu^{-1}$, rather than the behaviour we observe.

For completeness, and in order to add insight into the effects of Faraday inhomogeneity, we calculate the  Stokes parameters $Q$ and $U$ in the Faraday dominant limit as 
\be
U\approx U_o\cos{(\kappa_F L)},
\ee
and
\be
Q=-U_o\sin{(\kappa_FL)}.
\ee
The only way to suppress a large $U_o$ is to imagine many cells with random properties. Then $L$ is replaced by a characteristic $s$ for each cell and, in the mean,
 the linear polarization is suppressed \cite{BF2002}. The same suppression would occur in a depolarizing sheath.
\newpage

\section{X-ray Emission via Inverse Compton}
\label{appendixD}

Appendix~\ref{appendixA} outlines the jet model in which we imagine that the X-ray emission around the time of
the peak of the X-ray light curve,  
comes from
a small X-ray `lobe' (see Fig.~\ref{fig:cone}). We wish to link the currently observed radio emission to a time approximately
a year earlier 
to determine whether conditions at that time could account for the X-ray peak flux via 
inverse Compton upscattering,
also known as Synchrotron Self-Compton, or SSC. 
As before, our calculations are order-of-magnitude to check
for feasibility.

As indicated in the text (Eqn.~\ref{eq:gammaelec}), the critical frequency, $\nu_c$ (Hz), for a relativistic electron of
Lorentz factor, $\gamma_e$, in a magnetic field of strength, $B$,
is
\begin{equation}
\nu_c\,=\,4.2\,\times\,10^4\,{\gamma_e}^2\,B_\perp(-2)\label{eqn:critfreq}
\end{equation}
where $B_\perp(-2)$ is the perpendicular magnetic field strength
($B_\perp\,=\,B\,sin\,\theta$,  with $\theta$ the pitch angle). Hence for an electron to radiate
in the radio regime at 5 GHz (the
peak of the spectrum at time T1, Table~\ref{table:spectral_fits}), in a magnetic field of $B_\perp(-2) = 1$, a Lorentz
factor of $\gamma_e\,\approx\,347$ is required. Such a value could describe conditions in the radio
source around the time of the radio observations.

We need to consider the earlier time in the X-ray lobe at which $\gamma_e$ and $B_\perp(-2)$ are both
higher and the source size is smaller.
The SSC requirement is $\nu_X\,=\,{\gamma_e}^2\,\nu_c$, so from the above equation,
\begin{equation}
\label{eqn:xfreqssc}
\nu_X\,=\,4.2\,\times\,10^4\,{\gamma_e}^4\,B_\perp(-2)
\end{equation} 

The hard X-ray light curve of \cite{NW13} applies to the energy range, 17.3 - 80 keV and we adopt 30 keV, or
$\nu_X\,=\,7.2\,\times\,10^{18}$ Hz as the X-ray frequency.  Also, 
for a jet of constant velocity,
 $B_r\propto r^{-2}$ (see Appendix~\ref{appendixA}), so for an X-ray lobe that is a factor of 10 smaller than the
current radio source size ($10^{16}$ rather than $10^{17}$ cm, or
0.1 mas rather than 1 mas), the magnetic field strength can be estimated
to be
$B_\perp(-2)\,=\,100$ at the earlier time.  We can now find the Lorentz factor for relativistic electrons in the
X-ray lobe, which from Eqn.~\ref{eqn:xfreqssc}, is $\gamma_e\,=\,1109$. Then, by Eqn.~\ref{eqn:critfreq},
the synchrotron critical frequency would be $\nu_c\,=\,5.5\,\times\,10^{12}$ Hz or $\lambda_c\,=\,55~\mu$m.
Thus the synchrotron emission in the X-ray lobe would be in the infra-red at the earlier time.
Though not required for this development, one can show via standard formulae, that such emission would be optically thin.

We now need to calculate the infra-red flux in the X-ray lobe, given these conditions.  First, we find the
 specific intensity of the synchrotron emission, specifying infra-red with the subscript, IR,
\begin{eqnarray}
\label{eqn:synch_IR}
I_{\nu_{IR}}&=&\epsilon_{\nu_{IR}}\,s_X\\\nonumber
&=& C_5(p)\,N_{0_{IR}}\,{B_{IR}}^{(p+1)/2}\,\left(\nu_{IR}/2C_1\right)^{(1-p)/2}\left(sin(\theta_{IR})\right)^{(1+p)/2}\,s_X
\end{eqnarray}
where $\epsilon_{\nu_{IR}}$ is the emissivity 
 and $s_X$ is the line-of-sight distance through the X-ray lobe. Here, $C_5(p)$ and $C_1$ are Pacholczyk's constants
\citep{Pach70} for electron energy spectral index, $p$, and $N_{0_{IR}}$ is a normalizing factor for the energy spectrum.
(cf. Sect.~\ref{sec:adiabatic}).  We have explicitly expressed the magnetic field, $B_{IR}$ separated from 
 $sin\theta_{IR}$ because, in the next step, we will compare current radio values to $IR$ values at the earlier
time and, when
$\gamma_e$ changes, so does the pitch angle.

We now multiply and divide Eqn.~\ref{eqn:synch_IR} by the same expression for the (current) radio emission, 
$I{\nu_{R}}$, where $R$ refers to the radio, finding after
some manipulation,
\begin{eqnarray}
\label{eqn:firstratio}
I_{\nu_{IR}}&=&I{\nu_{R}}\,
\left(\frac{{N_0}_{IR}}{{N_0}_{R}}\right)
\left(\frac{{B}_{IR}}{{B}_{R}}\right)^{(1+p)/2}
\left(\frac{sin(\theta_{IR})}{sin(\theta_{R})}\right)^{(1+p)/2}
\left(\frac{\nu_{IR}}{\nu_{R}}\right)^{(1-p)/2}
\left(\frac{s_X}{s_R}\right)
\end{eqnarray}

A conversion to flux, $S_{IR}$, requires multiplying by the source solid angle and by the frequency.  
If we again take $p\,=\,2$
which is implied from the frequency spectral index in the radio, then Eqn.~\ref{eqn:firstratio} becomes

\begin{eqnarray}
\label{eqn:secondratio}
S_{IR}&=&S_{R}\,
\left(\frac{{N_0}_{IR}}{{N_0}_{R}}\right)
\left(\frac{{B}_{IR}}{{B}_{R}}\right)^{3/2}
\left(\frac{sin(\theta_{IR})}{sin(\theta_{R})}\right)^{3/2}
\left(\frac{\nu_{IR}}{\nu_{R}}\right)^{-1/2}
\left(\frac{s_X}{s_R}\right)
\end{eqnarray}

We can now relate the radio quantities, which are mostly known, to the infra-red quantities at the
earlier time to estimate $S_{IR}$.
We take $S_R\,=\,\nu_R\,S_{\nu_R}$, where  $\nu_R\,=\,5\,\times\,10^9$ Hz and 
$S_{\nu_R}\,=\,433$ mJy
(Table~\ref{table:spectral_fits} for time T1) to find  $S_R\,=\,2.1\,\times\,10^{-14}$ erg s$^{-1}$ cm$^{-2}$.
For an adiabatically expanding source, we found (Sect.~\ref{sec:adiabatic}), that
${{N_0}_{IR}}/{{N_0}_{R}}\,\propto\,(t_{IR}/t_{R})^{-2/3}$, where $t_R$, $t_{IR}$ are the times for the radio emission
and IR emission, respectively.  We can take $t_{IR}/t_{R}\,\approx\,1/2$, since the time from the outburst to the
radio observations is about twice the time from the outburst to the X-ray peak, leading to
${{N_0}_{IR}}/{{N_0}_{R}}\,\approx\,1.6$. The ratio, $B_{IR}/{B}_{R}\,=\,100$, from above. For the pitch angle, we
take $sin(\theta_{IR})/sin(\theta_{R})\,=\,{\gamma_e}_R/{\gamma_e}_{IR} = 347/1109 = 0.31$. The ratio of frequencies is
$\nu_{IR}/\nu_{R} = (5.5\,\times\,10^{12})/(5\,\times\,10^9) = 1.1\,\times\,10^3$ and the ratio of sizes is
${s_X}/{s_R} = 0.1$.

The result is $S_{IR}\,=\,1.9\,\times\,10^{-11}$ erg s$^{-1}$ cm$^{-2}$.

It remains to compute the X-ray flux from SSC.  From Eqns.~\ref{eqn:urad} and \ref{eqn:ssc} expressed
in a more convenient form 
in terms of the infra-red synchrotron emission in the shocked region that emits the
X-rays, 
\begin{equation}
S_{X_{SSC}}\,=\,S_{IR}\,\frac{U_{rad_{IR}}}{U_{mag}}\,=\,5.7\,\times\,10^{12}\,\frac{{S_{IR}}^2}{{\theta_s}^2\,{B_{IR}(-2)}^2}
\end{equation}
for $S_{X_{SSC}},~S_{IR}$ in erg s$^{-1}$ cm$^{-2}$, $\theta_s$ in mas and $B_{IR}(-2)$ in units of $10^{-2}$ G.
With the above value of $S_{IR}$, $\theta_s\,=\,0.1$ and $B_{IR}(-2) = 100$ (see above), we find,
$U_{rad_{IR}}/U_{mag}\,=\,1.1$ and
$S_{{X}_{SSC}}\,=\,2.0\,\times\,10^{-11}$ erg s$^{-1}$ cm$^{-2}$.  The peak of the X-ray light curve is
$S_{X}\,\approx\,8.0\,\times\,10^{-11}$ erg s$^{-1}$ cm$^{-2}$.

Thus, our estimated X-ray flux from SSC falls short of the observed X-ray peak by only a factor of 4.  The result is
quite sensitive to the choice of magnetic field in the radio which was taken as  $B_R(-2)\,=\,1$ above.  
For example, if $B_R(-2)\,=\,4$ then 
$U_{rad_{IR}}/U_{mag}\,=\,0.06$ and 
$S_{{X}_{SSC}}\,=\,1.0\,\times\,10^{-12}$ erg s$^{-1}$ cm$^{-2}$ which falls short of 
explaining the X-ray peak.  If $B_R(-2)\,=\,0.5$, then 
$S_{{X}_{SSC}}\,=\,1.0\,\times\,10^{-10}$ erg s$^{-1}$ cm$^{-2}$ which is more than adequate to explain the X-ray peak; however,
$U_{rad_{IR}}/U_{mag}\,=\,4.6$ which is uncomfortably large.

Since the result is so sensitive to the magnetic field, one needs to examine it more closely.  We have assumed that
$B_r\propto r^{-2}$
which accounts only for geometric expansion.  Yet, our jet model indicates that the X-ray lobe is
a shocked region that should enhance the magnetic field.  If we allow for geometric expansion, as before, but include a
modest enhancement to the magnetic field in the X-ray lobe of a factor of 1.8 (a strong shock would give factor of 4)
then  $U_{rad_{IR}}/U_{mag}\,=\,1.1$ and 
$S_{{X}_{SSC}}\,=\,8\,\times\,10^{-11}$ erg s$^{-1}$ cm$^{-2}$. With these conditions, 
SSC can account for the X-ray flux at the peak of the 
light curve.





\vskip 0.2truein

{\it Facilities:} \facility{VLA}.

\end{document}